\documentclass[12pt,reqno]{amsart}
\usepackage{graphicx,amssymb,amsmath}
\usepackage{amsaddr}
\usepackage[utf8]{inputenc}
\usepackage{epstopdf}
\usepackage{url,hyperref}	
\usepackage{bm}
\usepackage{color}
\usepackage{array}
\usepackage{enumerate}
\usepackage{booktabs}

\usepackage{theorem}
\usepackage[nodayofweek]{datetime}

\usepackage{natbib}
\setcitestyle{authoryear,open={(},close={)}}
\usepackage{hyperref}
\usepackage{etoolbox}
\usepackage{chemformula}
\usepackage{gensymb}

\newcommand{\eps}{{\varepsilon}}

\title{A model for Dansgaard-Oeschger events and millennial-scale abrupt climate change without external forcing}

\author[G. A. Gottwald]{Georg A. Gottwald}
\address[G. A. Gottwald]
{School of Mathematics and Statistics \\
 University of Sydney \\
 NSW 2006 \\
 Australia
}
\email[G. A. Gottwald]{georg.gottwald@sydney.edu.au}

\begin{document}

\maketitle


\begin{abstract}
We propose a conceptual model which generates abrupt climate changes akin to Dansgaard-Oeschger events. In the model these abrupt climate changes are not triggered by external perturbations but rather emerge in a dynamic self-consistent model through complex interactions of the ocean, the atmosphere and an intermittent process. The abrupt climate changes are caused in our model by intermittencies in the sea-ice cover. The ocean is represented by a Stommel two-box model, the atmosphere by a Lorenz-84 model and the sea-ice cover by a deterministic approximation of correlated additive and multiplicative noise (CAM) process. The key dynamical ingredients of the model are given by stochastic limits of deterministic multi-scale systems and recent results in deterministic homogenisation theory. The deterministic model reproduces statistical features of actual ice-core data such as non-Gaussian $\alpha$-stable behaviour. \\ The proposed mechanism for abrupt millenial-scale climate change only relies on the existence of a quantity, which exhibits intermittent dynamics on an intermediate time scale. We consider as a particular mechanism intermittent sea-ice cover where the intermittency is generated by emergent atmospheric noise. However, other mechanisms such as freshwater influxes may also be formulated within the proposed framework.
\end{abstract}


\section{Introduction}

A remarkable signature of the climate of the past $100$ kyrs are the so called Dansgaard-Oeschger (DO) events \citep{DansgaardEtAl83}. These events occurred during the last glacial period and are characterised by abrupt warming within a few decades of $5$-$10$ degrees followed by more gradual cooling over more than $1$ kyr back to the stadial period with DO events recurring on a millennial time scale \citep{GrootesStuiver97,YiouEtAl97,DitlevsenEtAl05}. They were first detected in time series of temperature proxies such as $\ch{O}^{18}$-isotopes concentrations in ice-cores collated in Greenland \Citep{GRIP,NGRIP}. The analysis of the ice-core data conveyed certain statistical features of DO events such that the abrupt warming events are consistent with non-Gaussian L\'evy jump processes (so called $\alpha$-stable processes) \citep{FuhrerEtAl93,Ditlevsen99}. The dynamic mechanism which gave rise to these events is still under debate. There exists a plethora of theories aimed at explaining their occurrence, ranging from conceptual models  to simulations of complex coupled atmosphere-ocean general circulation models (see the excellent reviews by \cite{Crucifix12} and by \cite{LiBorn19}). Most theories are built around the premise that the ocean is the main agent controlling the DO events, and that the ocean's meridional overturning circulation (MOC) is reduced by freshwater influx 
\citep{ManabeStouffer99,FriedrichEtAl10}. This hypothesis has been tested in ocean general circulation models by studying the ocean response to prescribed freshwater flushes \citep{WeaverHughes94,GanopolskiRahmstorff01,HaarsmaEtAl01,MeissnerEtAl08,TimmermannEtAl03}. How these freshwater fluxes were produced in the first place is, however, left out in these studies. There is a need to develop a self-consistent mechanism for DO events, which does not rely on external factors not covered by the model. Moreover, the pivotal role of internal ocean dynamics has been questioned by \citet{Wunsch06}. Therein it is argued that the ocean's net meridional heat transport is not sufficiently strong to cause the massive changes in temperature as suggested from the ice-core data, and that ``the oceanic tail may not necessarily be wagging the meteorological dog''. It has instead been recognised recently that DO events involve an intimate and complex interaction between the ocean, sea-ice and the atmosphere (see the comprehensive review by \cite{LiBorn19}). In particular the role of stochastic wind forcing \citep{MonahanEtAl08,DrijfhoutEtAl13,KleppinEtAl15}, the importance of sea-ice and its changes \citep{GildorTziperman03,LiEtAl05,PetersenEtAl13,DokkenEtAl13,ZhangEtAl14,KleppinEtAl15,HoffEtAl16,BoersEtAl18,SadatzkiEtAl19}, the vertical structure of the Nordic seas \citep{SinghEtAl14,JensenEtAl16} as well as inter-hemisphere coupling mediated by Southern Ocean winds \citep{BanderasEtAl12,BanderasEtAl14} have all been found to have a significant effect on the phenomenon of DO events. \\

Building on these current developments in our understanding of DO events, we develop here a conceptual model for millennial-scale abrupt climate change consisting of a coupled dynamical system modelling the interactions between the ocean, sea-ice and the atmosphere, without any external forcing such as prescribed freshwater fluxes. We do so in an entirely deterministic fashion. The importance of stochastic atmospheric dynamics \citep{MonahanEtAl08,DrijfhoutEtAl13,DokkenEtAl13,KleppinEtAl15} and the observed effective $\alpha$-stable statistics of the ocean temperature \citep{Ditlevsen99} are accounted for via deterministically self-generated noise in a multi-scale setting. On the slow scale the ocean is modelled by a Stommel two-box model \citep{Stommel61} which is forced by an intermittent sea-ice model on an intermediate time scale. The atmosphere enters the model in form of a Lorenz-84 model on the fastest time scale, modelling jet streams and baroclinic eddy activity \citep{Lorenz84}. We consider here the possibility of two atmospheric Lorenz-84 models, one for the Northern hemisphere and one for the Southern hemisphere \citep{BanderasEtAl12,BanderasEtAl14}. The strongly chaotic atmosphere gives rise to Gaussian noise on the slower time scales of the sea-ice and of the ocean. The crucial premise of our model is that sea-ice is intermittent and that its dynamics is punctuated by sporadic events of extreme large sea-ice cover. The effect of atmospheric forcing on the variations of sea-ice has long been recognised \cite{FangWallace94,VenegasMytak00,DeserEtAl02}. In our model chaotic weather dynamics deterministically generates intermittent sea-ice dynamics. The emerging weakly chaotic intermittent sea-ice dynamics then subsequently generates the necessary non-Gaussian L\'evy noise in the slow ocean dynamics, driving the ocean temperature abruptly from its glacial steady (noisy) state to a warmer unstable state.   

From a dynamical systems point of view the theoretical backbone of the model consists of statistical limit laws to generate stochastic processes by appropriately integrating deterministic chaotic dynamics and hinges on recent advances in the study of diffusive limits of deterministic multi-scale systems \citep{MelbourneStuart11,GottwaldMelbourne13c,KellyMelbourne16,ChevyrevEtAl19}. Therein it is shown that noise can be deterministically generated within a multi-scale system. If the driving fast process is {\em{strongly}} chaotic, the slow dynamics is, in the limit of infinite time-scale separation, in effect a stochastic differential equation driven by Brownian (possibly multiplicative) noise. The mechanism can be motivated heuristically as follow: within one slow time unit the slow dynamics integrates the chaotic fast process and, invoking a central limit type argument, one ends up with an effective Gaussian noise. However, as was shown by \citet{Ditlevsen99}, ice-core data exhibit a strong degree of non-Gaussian $\alpha$-stable dynamics. Anomalous $\alpha$-stable noise, or a L\'evy process, is characterised by jumps at all scales with non-zero probability of large jumps (see, for example, \cite{CheckinEtAl08} for an exposition of $\alpha$-stable processes). As for the Gaussian noise discussed above, $\alpha$-stable L\'evy noise can be deterministically generated in an entirely deterministic fashion. The deterministic origin of anomalous diffusion can be linked to intermittent fast dynamics in which the dynamics spends long temporal intervals near a marginally stable fixed point or periodic orbit before experiencing chaotic bursts \citep{GaspardWang88}. The central limit theorem which generated the Gaussian noise in the case of strongly chaotic non-intermittent dynamics ceases to be valid but can be replaced by a modified statistical law \citep{Gouezel04}. \cite{GottwaldMelbourne13c,ChevyrevEtAl19} showed that for multi-scale systems with a weakly chaotic intermittent fast driving process the limiting stochastic process of the slow dynamics is given by (possibly multiplicative) $\alpha$-stable noise\footnote{See \cite{GottwaldMelbourne13} for a definition of what constitutes strong and weak chaos.} We consider here intermittent sea-ice dynamics modelled by correlated additive and multiplicative noise (CAM) \citep{SuraSardeshmukh08,SardeshmukhSura09,PenlandSardeshmukh12,SardeshmukhPenland15}. CAM noise naturally arises in deterministic multi-scale systems for the effective slow dynamics \citep{SardeshmukhSura09,MajdaEtAl09}. Using statistical limit laws developed by \cite{KellerKuske00}, \cite{ThompsonEtAl17} showed that fast intermittent CAM noise can be used to generate $\alpha$-stable processes. Within the framework of statistical limit laws we can now highlight the dynamic function of the geophysical ingredients of our coupled ocean-atmosphere sea-ice model: using the classical central limit theorem, a fast atmospheric model generates intermittent Brownian CAM noise of the sea-ice dynamics on an intermediate time scale. The sea-ice dynamics then generates $\alpha$-stable noise on the slow oceanic time scale by means of a generalised central limit theorem. We show that the emerging stochastic dynamics of this coupled ocean-atmosphere and sea-ice model is able to generate abrupt changes in the temperature akin of DO events.\\

The paper is organised as follow. In Section~\ref{sec:Data} we perform an analysis of ice-core data confirming that the data are consistent with a dynamic process involving $\alpha$-stable noise. Section~\ref{sec:homo} provides a heuristic approach to deterministic generation of stochastic processes, covering both the Gaussian and the $\alpha$-stable case. Sections~\ref{sec:model} and ~\ref{sec:numerics} are the heart of the paper. Section~\ref{sec:model} introduces the deterministic coupled ocean-atmosphere and sea-ice model. Section~\ref{sec:numerics} provides numerical simulations illustrating the capability of the model to capture abrupt climate changes akin to DO events. We conclude in Section~\ref{sec:discussion} with a discussion.

%





\section{Time series analysis of ice-core data}
\label{sec:Data}
\begin{figure}[htbp]
\centering
\includegraphics[width = 0.98\columnwidth]{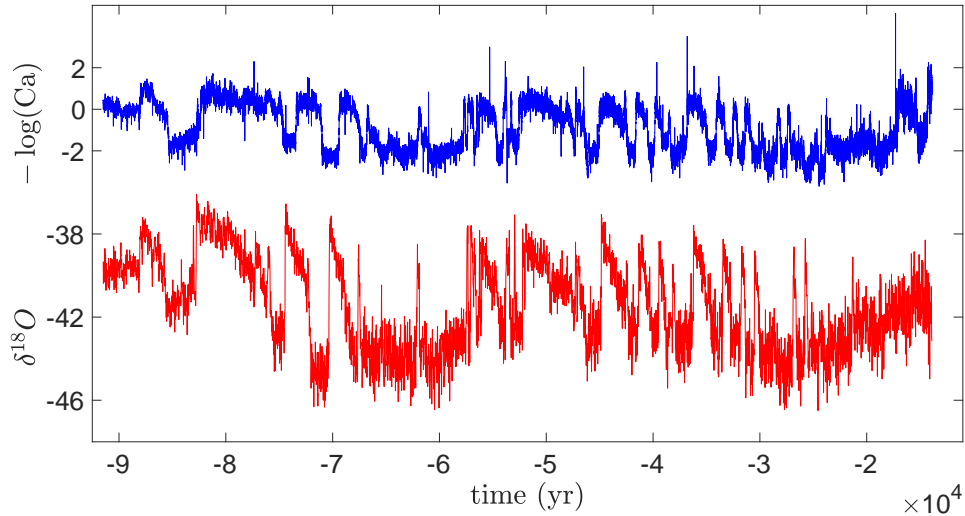}
\caption{The negative logarithm of the calcium concentration and $\delta^{18}\ch{O}$ as a function of time. The $\ch{Ca}$ time series was obtained from the GRIP ice-core data and have a temporal resolution of approximately $1$ year, and there are a total of $79,957$ data points between $11$ kyrs and $91$ kyrs. The $\delta^{18}\ch{O}$ time series was obtained from NGRIP ice-core data and have a temporal resolution of $20$ years with $6,114$ data points.}
\label{fig.data}
\end{figure}

Ice core data have immensely increased our knowledge about past climate variations \citep{GRIP,NGRIP}. An analysis of calcium ice core data collated in central Greenland as part of the GRIP programme \citep{FuhrerEtAl93} was performed by \cite{Ditlevsen99}. Calcium originates from dust deposited on the ice and is not diffusing as much as the usual $\delta^{18}\ch{O}$ proxy allowing for a higher temporal resolution. The logarithm of the calcium concentration is negatively correlated with $\delta^{18}\ch{O}$, with higher concentrations of $\ch{Ca}^{2+}$ in colder conditions due to enhanced exposure to sea shelves caused by lower sea levels, increased aridity and stronger zonal winds caused by an increased meridional temperature gradient \citep{FuhrerEtAl93,SchuepbachEtAl18}. The time series of $-\log(\ch{Ca})$ is shown in Figure~\ref{fig.data} together with the time series of $\delta^{18}\ch{O}$ illustrating their strong correlation. The data for $\delta^{18}\ch{O}$ were obtained from the NGRIP programme using the Greenland Ice Core Chronology (GICC05) time scale and the GICCO05modelext time scale for times past $60$kyr before year $2000$ \citep{VintherEtAl06,RasmussenEtAl06,AndersenEtAl06,SvenssonEtAl08,WolffEtAl10}. The time series of $\log(\ch{Ca})$ exhibits strong non-Gaussian character. \cite{Ditlevsen99} found that the data contain a significant $\alpha$-stable component with a stability parameter $\alpha=1.75$ in conjunction with multiplicative Gaussian noise.\\

We briefly revisit the analysis, using a different method to detect the $\alpha$-stable component. We assume that the data can be modelled by a one-dimensional stochastic differential equation of the form $dX = -U'(X)dt + \sigma_w dW_t + dL_\alpha$ where $W_t$ is standard Brownian motion and $L_\alpha$ is an $\alpha$-stable stochastic process. The prime denotes the derivative with respect to $X$. The potential $U(X)$ can be readily estimated from the data by using standard coarse graining of the data to estimate the conditional average of $dX$ \citep{Gardiner,Siegert98,Stemler07}. We obtain a quartic potential $U(X) = 0.0018\,X^4 - 0.0058\, X^3  +0.0024\, X^2 + 0.0028 \, X$ where the two potential well minima correspond to the stadial and interstadial regimes (see also \citep{KwasniokLohmann09,LohmannDitlevsen19}). The colder potential minimum is more stable than the warmer one. To estimate the presence of $\alpha$-stable noise we will not, as in \citet{Ditlevsen99}, study the scaling of the tails of the empirical probability density function (which scales as $X^{-\alpha-1}$), but rather employ the method of $p$-variation \citep{MagdziarzEtAl09,MagdziarzKlafter10,HeinImkellerPvalyukevich}. Whereas the presence of fat tails may also be caused by multiplicative Gaussian noise, $p$-variation is a proper statistics to isolate $\alpha$-stable behaviour. The statistics concerns the asymptotic behaviour of 
\begin{align*}
V_p^n(t) = \sum_{k=1}^{[n t]}|X(k/n) - X([k-1]/n)|^p.
\end{align*} 
This easily computable statistics measures the roughness of the process $X$, tuning into finer and finer partitions with increasing $n$. For $p=1$ the statistics reduces to the total variation and for $p=2$ it reduces to the quadratic variation. For Brownian motion where increments scale as $\sqrt{1/n}$ one obtains in the limit of $n\to \infty$ that $V_2^n(t) \sim t$, and $V_p^n(t)\to 0$ for $p>2$. Given an $\alpha$-stable process $X$ for some $\alpha<2$, the statistics $V_p^n(t)$ converges for $p>\alpha$ and diverges for $p<\alpha$. In \cite{HeinImkellerPvalyukevich} it was shown that if $X$ is a stochastic process $dX = -U'(X)dt + \sigma_w dW_t + dL_\alpha$ driven by $\alpha$-stable noise with $\alpha=p/2$ then $V_p^n(t)$ converges in distribution to $L_{1/2}$. This suggests to use a Kolmogorov-Smirnov test and find the value of $p=2\alpha$ for which the empirical cumulative distribution function is closest to the target cumulative distribution function of $L_{1/2}$. To estimate the cumulative distribution function we follow  \cite{HeinImkellerPvalyukevich} and choose to divide the $\ch{Ca}$ time series into $282$ segments, each consisting of $282$ data points. The minimal Kolmogorov-Smirnov distance is then found by varying the scale parameter of the target distribution $L_{1/2}$ for each value of $p$. The value $p^\star$ for which the minimum is attained then determines $\alpha=p^\star/2$. For details on the $p$-variation method see \citep{MagdziarzEtAl09,MagdziarzKlafter10,HeinImkellerPvalyukevich}. We remark that \cite{HeinImkellerPvalyukevich} found a value of $\alpha=0.75$, suggesting a L\'evy process with infinite mean. We find here, in reasonably close agreement with the result by \cite{Ditlevsen99}, the value of $\alpha=1.78$.  In our model, introduced in Section~\ref{sec:model}, the particular feature of DO events to exhibit $\alpha$-stable statistics will be generated by intermittent sea-ice dynamics. 

%


\section{Dynamic mechanism to generate Brownian motion and L\'evy noise from deterministic multi-scale systems}
\label{sec:homo}

The model developed in Section~\ref{sec:model} relies on recent developments in the study of stochastic limits of deterministic multi-scale systems The mathematical programme to derive limiting stochastic slow dynamics is coined homogenisation \citep{Givonetal04}. The machinery of homogenisation provides explicit expressions for the drift and diffusion components of the effective stochastic slow dynamics. In particular, we will use results from deterministic homogenisation of multi-scale systems \citep{MelbourneStuart11,GottwaldMelbourne13c,KellyMelbourne17,ChevyrevEtAl19}. Rather than stating the theorems we present here, following \cite{Gottwaldetal17}, a heuristic motivation to illustrate how deterministic multi-scale systems can give rise to an effective stochastic dynamics for the slow variables. Consider the slow-fast system for slow variables $x_\varepsilon$ and fast variables $y_\varepsilon$
\begin{align}
\dot x_\varepsilon & =\varepsilon^{\gamma-1}h(x_\varepsilon)f(y_\varepsilon), \quad x_\varepsilon(0)=x_0\label{e.ms0} \\
\dot y_\varepsilon & =\varepsilon^{-1}g(y_\varepsilon), \quad y_\varepsilon(0)=y_0,
\label{e.ms}
\end{align}
which is formulated  on the fast time scale. Here $\varepsilon\ll 1$ denotes the time scale separation and $\gamma\ge \tfrac12$. We assume that the fast dynamics is supported on a chaotic attractor and is statistically stationary in the sense that averages can be computed by means of temporal averages. Integration of the slow dynamics yields 
\begin{align*}
x_\varepsilon(t) 
& =x_0 + \varepsilon^{\gamma}\int_0^\frac{t}{\varepsilon} h(x_\eps(\tau))f(y_\eps(\tau))\,d\tau. 
\end{align*}
Introducing $n=\varepsilon^{-1}$ and $\alpha=1/\gamma$ we obtain
\begin{align}
x_\varepsilon(t) 
& = x_0 + \frac{1}{n^\frac{1}{\alpha}}\int_0^{tn}  h(x_\eps(\tau))f(y_\eps(\tau))\,d\tau .
\label{eq:xslow}
\end{align}
Consider first the case $\alpha=\gamma=1$, then for $n\to \infty$ (or equivalently for $\epsilon \to 0$) the temporal integral is simply the average over the fast dynamics, and by the law of large numbers (the most simple statistical limit law) the slow dynamics remains deterministic in the limit $\varepsilon \to 0$, and solutions $x_\varepsilon(t)$ converge to solutions of the deterministic equation $\dot X = F h(X)$ with $X(0)=x_0$ where $F\equiv {\rm{const}}$ is the average over the fast dynamics of $f(y_\eps)$. Now consider the case when the average is zero with $F\equiv 0$. Clearly, the implied deterministic limit $X(t)=X(0)$ does not capture the dynamics of the solution $x_\eps(t)$ of the actual multi-scale system which is constantly driven by non-zero $f(y_\eps(t))$. One needs to go to longer time scales to see these fluctuations sum up to generate noise. This can be seen from  (\ref{eq:xslow}) by setting $\alpha=2$ (i.e. $\gamma=\tfrac12$). For $\alpha=2$ the integral is reminiscent of the central limit theorem. Indeed using statistical limit laws for strongly chaotic dynamical systems \citep{MelbourneNicol05,MelbourneNicol09}, the integral term converges to Gaussian noise. For the purpose of this exposition it is sufficient to think of {\em{strongly}} chaotic dynamical systems as systems for which the auto-correlation function is integrable; this will be contrasted to {\em{weakly}} chaotic dynamical systems for which the auto-correlation function is not integrable \citep{GottwaldMelbourne13}. It is important to note that it is not the chaotic signal $y_\eps$ itself that is noisy but rather the integrated fast chaotic variable. Care has to be taken in what way the stochastic integral in (\ref{eq:xslow}) is to be interpreted \citep{GottwaldMelbourne13c,KellyMelbourne17}. In the case of $1$-dimensional slow variables $x_\eps$, which will be considered in Section~\ref{sec:model} for the sea-ice model, the stochastic integrals are in the sense of Stratonovich, i.e. classical calculus is preserved in the limiting process when passing from the smooth deterministic multi-scale system to the rough stochastic differential equation. In this case, the slow dynamics of the multi-scale system (\ref{e.ms0})--(\ref{e.ms}) converges on the slow times $x_\eps(t/\eps)\to X(t)$ where $X$ satisfies the stochastic differential equation
\begin{align}
dX = \Sigma \, h(X)\circ dW_t,
\end{align}
with standard Brownian motion $W_t$ (and $\circ$ denoting that the noise is to be interpreted in the sense of Stratonovich) and the diffusion coefficient is given by the Green-Kubo formula
\begin{align*}
\Sigma = \int_0^\infty C(t)\, dt,
\end{align*}
with normalised auto-correlation
\begin{align*}
C(t)=\frac{1}{\sigma^2}\int_0^\infty f_0(y_\eps(t+s))f_0(y_\eps(s))\, ds
\end{align*}
with $C(0)=1$. The diffusion coefficient $\Sigma$ is well defined if the auto-correlation function is integrable.

There is, however, a class of weakly chaotic dynamical systems, for which the central limit theorem breaks down and fluctuations are of the L\'evy type rather than Gaussian. Weakly chaotic dynamics is characterised by intermittent behaviour where the dynamics spends extensive time near ``sticky" equilibria or periodic orbits before sporadic excursive bursts away from those marginally unstable objects. It has recently been shown that, if $f(y_\eps)$ is non-zero in the laminar phase, the central limit theorem can be replaced for weakly chaotic dynamics and the integral term in (\ref{eq:xslow}) converges in distribution to a stable law $L_{\alpha,\eta,\beta}$ of exponent $\alpha\in(1,2)$ \citep{Gouezel04}. The stability parameter $\alpha$ determines the algebraic decay in the tail of the probability density function, the scale parameter $\eta$ measures the spread of the distribution around its maximum and the skewness parameter $\beta$ encapsulates the probability of the process experiencing a positive jump or negative jump with $\beta=\pm 1$ having only positive/negative jumps. \cite{GottwaldMelbourne13c,ChevyrevEtAl19} showed that for intermittent fast dynamics (\ref{e.ms}) solutions $x_\varepsilon$ converge weakly to solutions of the stochastic differential equation
\begin{align} 
\label{eq-Marcus}
dX=h(X)\diamond dL_{\alpha,\eta,\beta}, \quad X(0)=x_0.
\end{align}
The parameters $\alpha$, $\beta$ and $\eta$ of the L\'evy process $L_{\alpha,\eta,\beta}$ are determined by the statistical properties of the driver $f(y_\eps)$. The diamond denotes that the noise $h(X)\diamond dL$ is to be interpreted in the sense of Marcus \citep{Marcus81,Applebaum,ChechkinPavlyukevich14}. The Marcus interpretation is the analogue of the Stratonovich interpretation for Brownian noise in the sense that classical calculus prevails, consistent with the intuition that one expects that the noise arises as a limit involving only smooth functions of a smooth deterministic system, and hence classical calculus should be inherited by the limiting system. We remark that the noise is of Marcus type independent of the dimension of the slow variables, unlike for the Stratonovich interpretation in the case of Brownian motion which is only ensured for $1$-dimensional slow variables. The Marcus integral $\int^t h(X)\diamond dL_{\alpha,\eta,\beta}(s)$ involves cumbersome expressions such as sums over infinitely many jumps. Moreover, whereas one can pass readily between the Stratonovich integrals to It\^o integrals, this is not possible for Marcus integrals. In our applications here, however, the $\alpha$-stable noise will be additive and these issues do not arise. The convergence to a L\'evy process can be heuristically understood by realising that if the dynamics $y_\eps$ is near the marginally unstable fixed point $y_\eps=y_\eps^\star$, the slow dynamics is driven by a constant $h(x_\eps)f(y_\eps^\star)$ (note that on the fast time scale $\tau=t/\epsilon$ $x_\eps$ is approximately constant). Hence the slow variable experiences ballistic drift during the laminar phases. It is those long ballistic drifts which amount to the jumps of the L\'evy process when viewed on a long time scale (see \cite{GottwaldMelbourne13,GottwaldMelbourne16,GottwaldMelbourne20} for numerical illustrations of this mechanism).

In a different strand of work, based on statistical limit laws for stochastic dynamical systems \citep{KellerKuske00}, \cite{ThompsonEtAl17} argue that so called correlated additive and multiplicative (CAM) noise processes 
\begin{align}
dy_\eps = L y_\eps \, dt - \frac{E}{2} G\, dt + (E y_\eps+G) \circ dW_1 + B \,dW_2
\label{eq:CAM}
\end{align}
with independent standard Brownian motions $W_{1,2}$ and $L<0$ lie in the domain of $\alpha$-stable processes which means that they give rise to $\alpha$-stable processes when integrated. For $B\neq 0$ the mean is well defined and one has explicit expressions for the parameters of the resulting L\'evy process $\alpha$, $\beta$ and $\eta$ as functions of the parameters of the CAM process \citep{KellerKuske00,ThompsonEtAl17}. The stability parameter $\alpha$ of the resulting $\alpha$-stable process $L_{\alpha,\eta,\beta}$ is given by 
\begin{align}
\alpha&=-2\frac{L}{E^2},
\label{eq:alphaCAM}
\end{align}
the skewness parameter is given by 
\begin{align}
\beta=\tanh(\frac{\pi G(\alpha-1)}{2B})
\label{eq:betaCAM}
\end{align}
and the scale parameter $\eta$ is given by 
\begin{align*}
\eta=\left(\frac{2\cosh(\frac{\pi G(\alpha-1)}{2B})}{E^{\alpha+1}\alpha N} \Gamma(1-\alpha)\cos(\frac{\pi}{2}\alpha)\right)^\frac{1}{\alpha}
\end{align*}
with
\begin{align*} 
N&=2\pi(2B)^{-\alpha}\frac{\Gamma(\alpha)}{E\Gamma(z)\Gamma(\bar z)},\qquad \quad
z=\frac{\alpha+1}{2}+ i\frac{G(\alpha-1)}{B},
\end{align*}
where the bar denotes the complex conjugate. 

\begin{figure}
	\centering
	\includegraphics[width=0.47\linewidth]{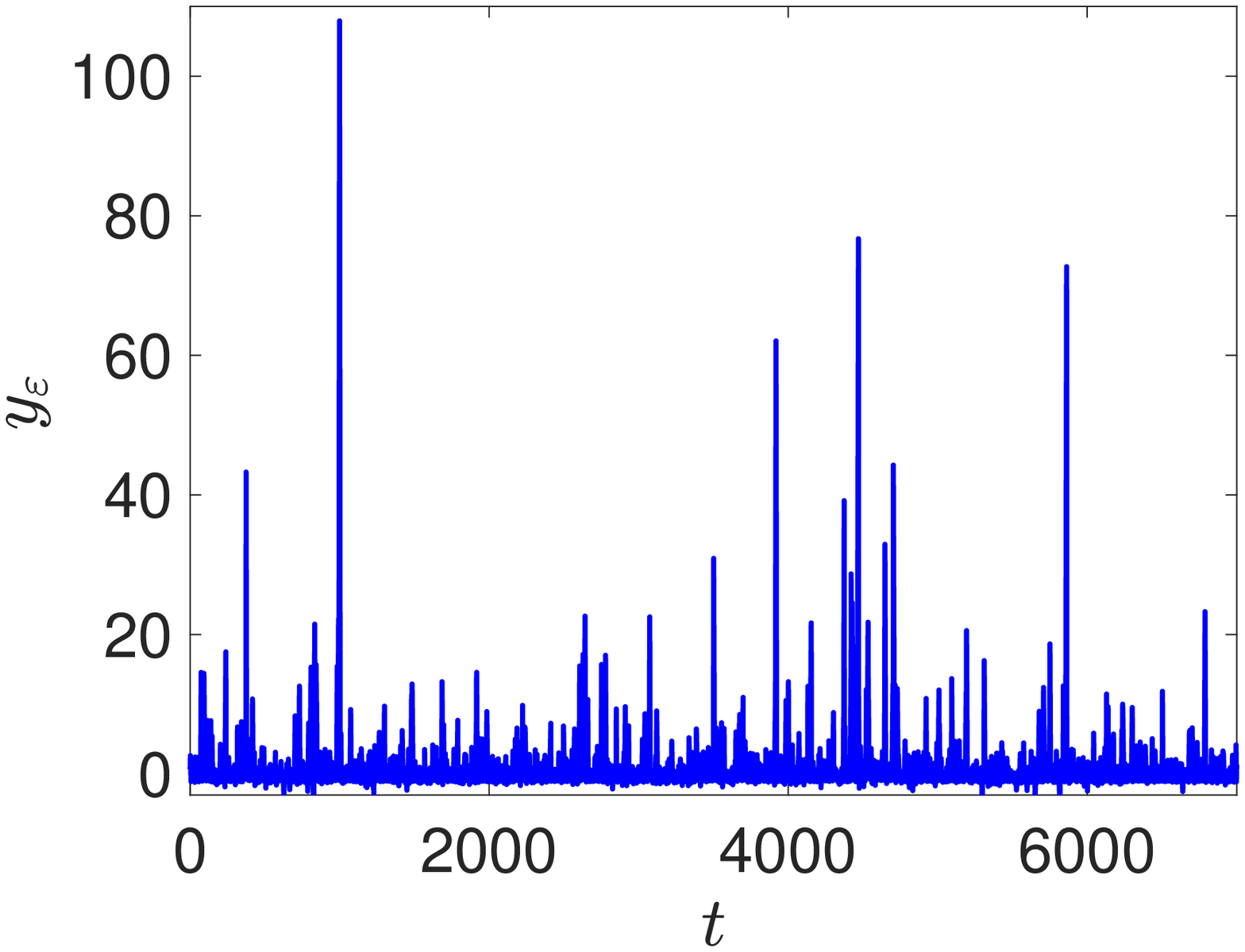}		
	\hspace{1mm}
	\includegraphics[width=0.47\linewidth]{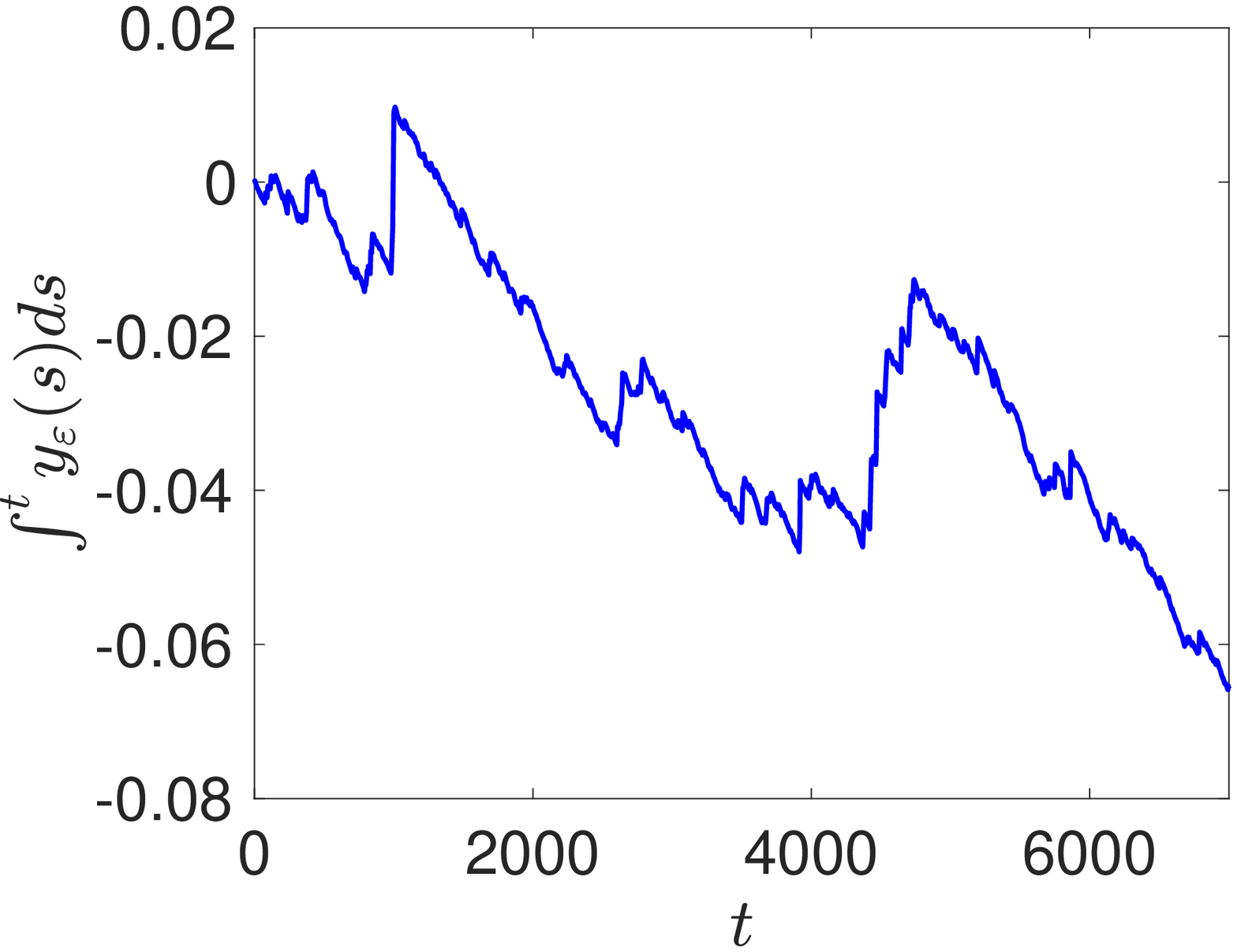}	
	\caption{Left: Realisation of a CAM process with $(L,E,G,B)=(-094,1.118,1,0.3)$. Right: Approximation of an $\alpha$-stable process with $\alpha=1.5$ and $\beta=0.99$ from the time series shown on the left.}
	\label{fig:CAM}
\end{figure}

Figure~\ref{fig:CAM} shows an example of a time series of a CAM process with $L=-0.94$, $E=1.118$ $G=1$ and $B=0.3$, implying that $\xi = \eps^\gamma\int^{t/\eps}f(y_\eps(s))ds$ with $\gamma=1/\alpha$ converges to an $\alpha$-stable process with $\alpha=1.5$ and $\beta=0.99$ (implying that there are almost only upwards jumps). Here the mechanism of generating $\alpha$-stable noise is different to the one described above: rather than the jumps consisting of many small jumps during the long laminar phases of varying length, the jumps here are caused by the sporadic peaks of varying sizes.\\  

In Section~\ref{sec:model} we shall model sea-ice by a deterministic approximation of a CAM process, whereby the two independent Brownian motions $W_{1,2}$ are approximated by two uncorrelated fast strongly chaotic processes, along the lines described above.


\section{Coupled ocean-atmosphere and sea-ice model}
\label{sec:model}

\begin{figure}
	\centering
	\includegraphics[width=0.99\linewidth]{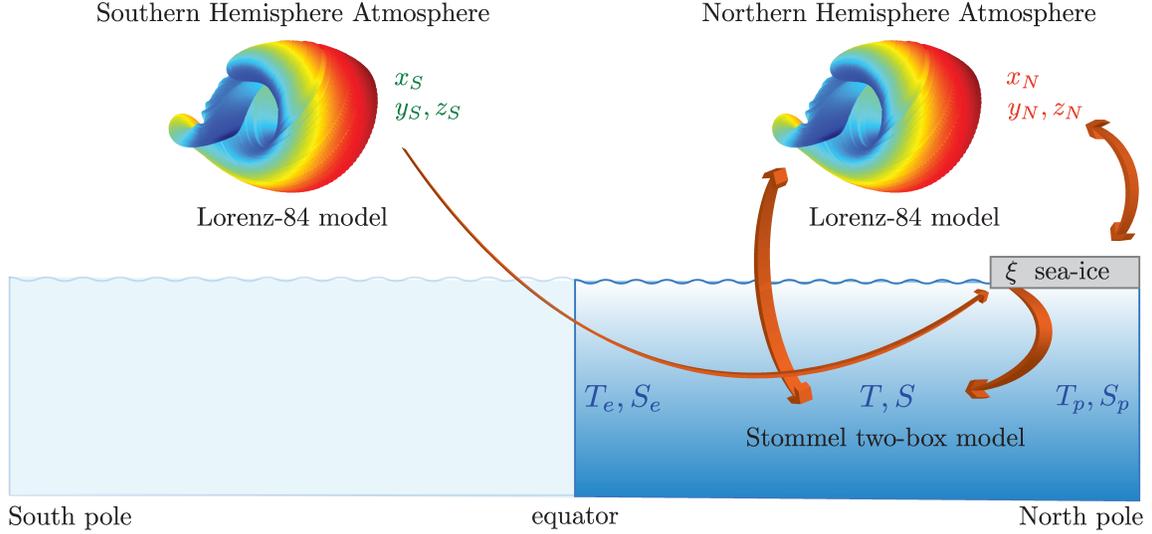}	
	\caption{Schematic of the coupled ocean-atmosphere and sea-ice model, highlighting the interdependencies and the characterising variables.}
	\label{fig:DOModel_sketch}
\end{figure}

We construct a conceptual deterministic coupled ocean-atmosphere and sea-ice model. The ocean model is given by a Stommel two-box model \citep{Stommel61} and the atmosphere is represented by a Lorenz-84 model, decsribing the westerly jet stream and large-scale eddies \citep{Lorenz84}. The sea-ice is modelled by a linear intermittent CAM process driven by the fast atmosphere and is characterised by sporadic brief periods of large sea-ice extent (cf. Figure~\ref{fig:CAM}). The intermittent character of the sea-ice is the main premise of our model and is paramount to generate the abrupt climate changes of DO events. The abrupt climate changes are a signature of an emerging $\alpha$-stable driving signal induced by integrated intermittent sea-ice dynamics. To deterministically generate the $\alpha$-stable noise on the slow oceanic time scale using the statistical limit theorems outlined in Section~\ref{sec:homo}, two further scales are required besides the slow oceanic time scale: a fast and an intermediate time scale. The fast strongly chaotic atmosphere dynamics integrates on the intermediate time scale of the sea-ice to Brownian motion to generate CAM noise. Then the CAM noise is integrated on the slow oceanic time scale to generate $\alpha$-stable L\'evy noise. We impose the natural time scale separation of the slow ocean with the typical diffusive time scale estimated as  $219$ years \citep{Cessi94}, an intermediate sea-ice dynamics occurring on time scales of months and a fast atmosphere with typical time scales of days. This suggest to introduce time scale parameters for the fast atmosphere $\epsilon_f$ and the intermediate sea-ice dynamics $\epsilon_i$ as 
\begin{align}
\epsilon_f &= \frac{1}{365\times 219}\approx 1.25 \times 10^{-5},\\
\epsilon_i &= \frac{30}{365\times 219}\approx 3.75 \times 10^{-4}.
\label{eq:eps}
\end{align} 
The ocean is characterised by coarse meridional temperature and salinity gradients
\begin{align}
T&=T_e-T_p,\\
S&=S_e-S_p,
\end{align}
where the subscripts $e$ and $p$ denote the respective values at equatorial and polar locations. The sea-ice dynamics is characterised by the extent of the sea-ice cover $\xi$. The atmosphere is characterised by the westerly zonal mean flow $x_{N,S}$ and the superimposed large scale eddies with amplitudes $y_{N,S}$ and $z_{N,S}$. Subscripts $N$ and $H$ denote the respective values of the Northern and Southern hemisphere. We first present the coupled non-dimensional model (\ref{eq:ocean})--(\ref{eq:atmosphere}) for these variables together with the coupling terms (\ref{eq:FG})--(\ref{eq:sigma}) capturing the various interactions between the ocean, atmosphere and sea-ice, before deriving the model and the non-standard coupling terms in Sections~\ref{sec:41}--\ref{sec:43}. Figure~\ref{fig:DOModel_sketch} presents a schematic illustrating the model and its various dependencies. For ease of navigation relevant variables and parameters are listed in Table~\ref{tab:DO}.

\begin{table}[h]
\begin{center}
	\begin{tabular*}{0.98\textwidth}{l l l}
	\toprule
	\multicolumn{3}{c }{\textbf{fast atmosphere: Lorenz-84 model}}\\
	\multicolumn{3}{c }{\scriptsize{(for Northern (H) and Southern (S) hemisphere)}}\\
	\midrule
	$x_{N,S}$ & strength of westerly zonal mean flow\\
	$y_{N,S}$, $z_{N,S}$  & amplitude of sine and cosine phase of large-scale eddy\\
	$\Delta_{N,S}$ & eddy energy with $\Delta = y^2+z^2$ \\
	$F^{N,S}$ & meridional temperature gradient\\
	$G^{N,S}$ & longitudinal temperature gradient \\
	\midrule
	\multicolumn{3}{c }{\textbf{intermediate sea-ice model: CAM noise }} \\
	\midrule
	$\xi$ & sea-ice cover\\
	\midrule
	\multicolumn{3}{c }{\textbf{slow ocean: Stommel two-box model}}\\
	\midrule 
	$T$ & temperature gradient $T=T_e-T_p$ between\\ & equatorial and polar ocean\\
	$S$ & salinity gradient $S=S_e-S_p$ between\\& equatorial and polar ocean\\
	$\Theta$ & ambient temperature gradient\\
	$\sigma$ & freshwater flux\\
	\midrule
	\multicolumn{3}{c }{\textbf{global coupling parameters}}\\
	\midrule 
	$\epsilon_f$ & ratio of characteristic time scales of fast atmosphere and slow ocean\\
	$\eps_i$ & ratio of characteristic time scales of intermediate sea-ice and slow ocean\\
	$\gamma$ & inverse of stability parameter of the $\alpha$-stable process with $\gamma=1/\alpha$\\
	\bottomrule
	\end{tabular*}		
	\caption {Variables and parameters used for the coupled ocean-atmosphere and sea-ice model.}
	\label{tab:DO} 
\end{center}
\end{table}

Specifically, we propose the following model: the ocean is described by a Stommel two-box model
\begin{align}
\label{eq:ocean}
\dot T  &= - \frac{1}{\epsilon_a}\left(T-\Theta(t)\right) - T  - \mu |S-T| T  - \frac{1}{\epsilon_i^{1-\gamma}}d\,(\xi-\bar\xi) T\\
\dot S  &= \sigma(t) - S - \mu |S-T| S,
\label{eq:ocean_S}
\end{align}
where $\epsilon_a$ measures the relaxation of the ocean temperature to the ambient temperature $\Theta(t)$, $\mu$ quantifies the transport strength and $\sigma(t)$ denotes freshwater flux. A more detailed definition of the parameters is provided in Section~\ref{sec:41}. The parameter $\gamma$ controls the application of the statistical limit theorems discussed in Section~\ref{sec:homo} to generate L\'evy noise with stability parameter $\alpha=1/\gamma$. The ocean-dynamics couples to the sea-ice dynamics
\begin{align}
\epsilon_i \dot \xi &= (\lambda + \frac{\kappa^2}{2})\xi + \sqrt{\frac{\epsilon_i}{\epsilon_f}}\delta\, (\kappa\xi+g)(x_S-\bar x_S)
+ \sqrt{\frac{\epsilon_i}{\epsilon_f}}c\,  (\Delta_N-\bar\Delta_N), 
\label{eq:seaice}
\end{align}
where the sea-ice dynamics is driven by the Northern hemisphere atmosphere through the eddy strength $\Delta_N=y_N^2+z_N^2$ and by the Southern hemisphere atmosphere by the jet stream $x_S$. The parameters $\lambda,\kappa,\delta,g,c$ allow for tuning of the $\alpha$-stable noise emerging in the ocean model (\ref{eq:ocean}) (cf. (\ref{eq:CAM})). The atmospheres of the Northern and Southern hemisphere are modelled by two Lorenz-84 systems
\begin{align}
\epsilon_f \dot x_{N,S} &= -(y_{N,S}^2+z_{N,S}^2) - a^{(N,S)}\, (x_{N,S}-F^{(N,S)})\\
\epsilon_f \dot y_{N,S} &= x_{N,S}\, y_{N,S}-b^{(N,S)}\, x_{N,S}\, z_{N,S} - (y_{N,S}-G^{(N,S)})\\
\epsilon_f \dot z_{N,S} &=  b^{(N,S)}\, x_{N,S}\, y_{N,S}+x_{N,S}\, z_{N,S}- z_{N,S}.
\label{eq:atmosphere}
\end{align}
To generate Brownian motion in the sea-ice dynamics (\ref{eq:seaice}) the only requirement for the choice of the parameters $a^{(N,S)}$, $b^{(N,S)}$, $F^{(N,S)}$ and $G^{(N,S)}$ is that the Lorenz-84 systems supports chaotic dynamics. The southern meridional and longitudinal temperature gradients $F^{(S)}$ and $G^{(S)}$ are set to constant $F^{(S)}=F^{(S)}_0$ and $G^{(S)}=G^{(S)}_0$ whereas the northern meridional and longitudinal temperature gradients $F^{(N)}$ and $G^{(N)}$ include back-coupling to the ocean dynamics and the sea-ice via
\begin{align}
\label{eq:FG}
F^{(N)} &= F^{(N)}_0 + F^{(N)}_1T + F^{(N)}_2\xi  \\
G^{(N)}  &= G^{(N)}_0 - G^{(N)}_1 T - G^{(N)}_2\xi  ,
\end{align}
with $F^{(N)}_{1,2}\ge 0$ and $G^{(N)}_{1,2}\ge 0$. The ambient temperature gradient $\Theta(t)$ of the ocean is driven by the atmosphere via thermal wind balance and is modelled as
\begin{align}
\Theta(t) & = \theta_0 + \theta_1 \frac{x_N-\bar x_N}{\sqrt{\epsilon_f}} ,
\label{eq:Theta}
\end{align}
and the salinity gradient $S$ is driven by the freshwater flux $\sigma(t)$ which is affected by both the atmosphere and the sea-ice, and is modelled as 
\begin{align}
\sigma(t) &= \sigma_0 + \sigma_1 \frac{\Delta_N-\bar \Delta_N}{\sqrt{\epsilon_f}} + \sigma_2 \frac{\dot \xi - \bar{\dot \xi}} {\epsilon_i^{1-\gamma_\xi}}.
\label{eq:sigma}
\end{align}


The model (\ref{eq:ocean})--(\ref{eq:atmosphere}) includes a wide range of interactions  between the ocean, the atmosphere and the sea-ice, captured in (\ref{eq:FG})--(\ref{eq:sigma}). To obtain abrupt warming events, however, it is sufficient to consider a minimal model with $F_1^{(N)}=F_2^{(N)}=G_1^{(N)}=G_2^{(N)}=\theta_1=\sigma_1=\sigma_2\equiv 0$. To reproduce realistic stochastic variations, however, we include atmospheric noise on the ocean dynamics and allow for $\theta_1\neq 0$ and $\sigma_1\neq 0$ in the numerical simulations presented in Section~\ref{sec:numerics}. \\

We derive the model (\ref{eq:ocean})--(\ref{eq:atmosphere}) with its coupling terms (\ref{eq:FG})--(\ref{eq:sigma}) in the following subsections. We begin by first deriving the classical Stommel two-box model on the slow time scale. We then continue setting up the atmosphere dynamics on the fastest time scale with a Lorenz-84 model and discuss how the atmosphere and the ocean couple. Finally, we set out to propose our model for the intermittent sea-ice dynamics and discuss how it modifies the dynamics of the (northern) atmosphere and ocean.


\subsection{Ocean model}
\label{sec:41}
We first formulate the ocean model on the slow time scale. We consider here the Stommel two-box model for the temperatures $T_{e,p}$ and salinities $S_{e,p}$ of an equatorial ocean box and a polar ocean box, respectively, \citep{Stommel61}. Although the derivation is standard and the box model is part of the canonical suite of conceptual models we present the derivation to illustrate the order of magnitude of the respective parameters of our model. We follow here \cite{Cessi94} and \cite{Roebber95} in the derivation. From conservation of heat, salt and water mass one obtains
\begin{align*}
\dot T_e  &= -\frac{1}{t_r}\left( T_e - \Theta_e(t) \right)  -\frac{1}{2} \Psi(\Delta \rho)\left(T_e-T_p\right) \\
\dot T_p  &= -\frac{1}{t_r}\left( T_p + \Theta_p(t) \right) -\frac{1}{2} \Psi(\Delta \rho)\left(T_p-T_e\right) \\
\dot S_e  &= \frac{W_e(t)}{H} - \frac{1}{2} \Psi(\Delta \rho)\left(S_e-S_p\right) \\
\dot S_p  &= -\frac{W_p(t)}{H} - \frac{1}{2} \Psi(\Delta \rho)\left(S_p-S_e\right)  .
\end{align*}
Here $\Theta_{e,p}(t)$ are the ambient atmospheric temperatures the ocean would equilibrate to on a relaxation time $t_r$ without any mass and heat exchange.  The flux $\Psi(\Delta \rho)$, capturing the mass and heat exchange, is driven by the density difference $\Delta \rho = \rho_e-\rho_p$ between the two ocean boxes. The densities are assumed to be linearly related to the temperature and salinity with $\rho_{e,p}/\rho_0 = 1+\alpha_s(S_{e,p}-S_0)-\alpha_T(T_{e,p}-T_0)$. The functions $W_{e,p}$, scaled with the typical height of the boxes $H$,  model salinity sources or sinks $W_{\rm precept}$ associated with precipitation/evaporation and/or freshwater sources $W_{\rm fresh}$ stemming from melting land ice. (Note that with slight abuse of notation, we use $W$ in this section to denote the salinity sinks and sources, and use $W$ otherwise to denote Brownian motion). We set $W_e(t) = W_{\rm prec}(t)/2$ and $W_p(t) = W_{\rm prec}(t)/2 + W_{\rm fresh}(t)$.\\ Introducing the coarse meridional temperature and salinity gradients $T=T_e-T_p$ and $S=S_e-S_p$ we obtain
 \begin{align}
\dot T  &= -\frac{1}{t_r}\left(T-\Theta(t)\right) - \Psi(\Delta \rho) T 
\label{e.2Bd1} \\
\dot S  &= \frac{W_e(t)+W_p(t)}{H} - \Psi(\Delta \rho) S  ,
\label{e.2Bd2}
\end{align}
with $\Theta(t)=\Theta_e(t)-\Theta_p(t)$. Following \cite{Stommel61} the flux is assumed to involve a diffusive component on the diffusive time scale $t_d$ and a hydraulic component of a Poiseuille flow with transport coefficient $q$, and we write
\begin{align}
\Psi(\Delta \rho) &= \frac{1}{t_d} + \frac{q}{V}|\Delta \rho| \nonumber \\
                        & =  \frac{1}{t_d} + \frac{q\rho_0}{V}|\alpha_s S - \alpha_T T|  ,
\end{align}
where $V$ denotes the typical volume of the boxes.\\ 

The equations (\ref{e.2Bd1})--(\ref{e.2Bd2}) are non-dimensionalised by scaling time with the diffusive time $t_d$, temperature with a characteristic temperature $T^\star$ and salinity with $\alpha_T T^\star/\alpha_S$. Introducing $\epsilon_a=t_r/t_d$ we arrive at
\begin{align}
\dot T  &= -\frac{1}{\epsilon_a}\left(T-\Theta(t)\right) - T - \mu |S-T| T 
\label{e.2Bnd1}
\\
\dot S  &= \sigma(t) - S -\mu |S-T| S  .
\label{e.2Bnd2}
\end{align}
Here $\mu = t_d q \rho_0 T_0\alpha_T/V$ and $\sigma(t) =\alpha_S t_d (W_{\rm prec}(t)+W_{\rm fresh}(t))/(\alpha_T T^\star H)$. We refer to \citep{Cessi94,Roebber95} for typical parameters. Typical relaxation times are $t_r=25$  days for the relaxation of the ocean surface, $t_r=5$ years for relaxation at a depth of $400$ m, $t_r=10$ years for relaxation at a depth of $800$ and $t_r=75$ years for the relaxation of the deep ocean. If we use the relaxation time at a typical ocean depth of $400$ m, we estimate $t_r=5$ years, which yields $\epsilon_a=0.0228$. Depending on whether we choose the ocean surface, depths at $400$ m, $800$ m or the deep ocean we estimate $\epsilon_a=\{3\times 10^{-4}, \, 0.0228, \, 0.046, \, 0.34\}$. The results presented in Section~\ref{sec:numerics} are not sensitive to the choice of depth. The box model has a typical ocean depth of $H=4500$ m and the control volume is estimated as $V=HL\delta_w$ where the typical meridional  scale is $L=8,250$ km and the width of the western boundary current is roughly $\delta_w=300$ km. The typical density is $\rho_0=1,029$ kg $m^{-3}$. The reference temperature is chosen to be $T^\star=20^o$C, and $\alpha_T=0.17\times 10^{-3}\;{\rm{C}}^{-1}$ and $\alpha_S = 0.75\times  10^{-3}\;{\rm{psu}}^{-1}$. The flux parameter $\mu$ is the ratio between the advective time scale and the diffusive time scale with $\mu=t_{ad}/t_d$. The advective time scale is calculated as follows: The western boundary current transports $B=12\, {\rm{Sv}}=12\times 10^6 {\rm m}^3 s^{-1}$. The advective time scale is then $t_{ad} = HL\delta_w/B = 29.4$ years which yields $\mu=7.5$. The freshwater flux in the North Atlantic is estimated as $(W_{\rm prec}(t)+W_{\rm fresh})S_0 \approx 0.2\, {\rm{Sv}}$ with $S_0=35{\rm{pus}}$ \citep{GanopolskiRahmstorff02}. Hence $\sigma = 0.95$. The diffusive time-scale is estimated as $t_d=L^2/\pi^2 \kappa_H=219$ years, where $\kappa_H=1000$ $\rm{m}^2 s^{-1}$ is the horizontal diffusion coefficient. Since we scale with the diffusive time scale, one unit of time corresponds to $219$ years, which defines the slow ocean time scale.\\

\begin{figure}
	\centering
	\includegraphics[width=0.5\linewidth]{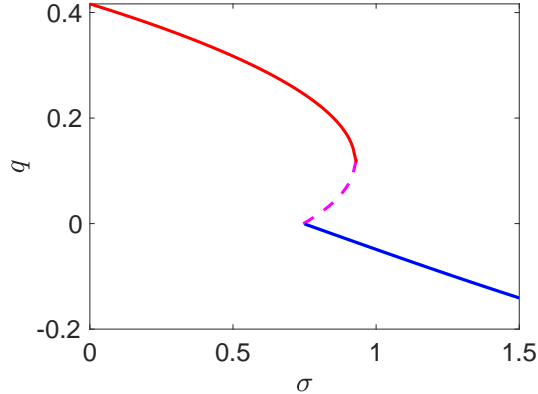}	
	\caption{Flow strength $q=T-S$ as a function of the freshwater flux $\sigma$ for $\mu=7.5$, $\Theta=1$ and $\epsilon_a=0.34$ for the Stommel box model (\ref{e.2Bnd1})--(\ref{e.2Bnd2}). The red branch depicts stable thermally driven steady states, the dashed curve depicts unstable solutions and the lower blue branch depicts salinity driven steady states. The Stommel box model exhibits bistability for $\sigma \in [0.75 0.94]$.}
	\label{fig:Stommel_q}
\end{figure}

The Stommel box model exhibits bistability for certain parameter ranges with one stable solution being thermally controlled with $q=T-S>0$ and the other controlled by salinity with $q<0$. Figure~\ref{fig:Stommel_q} shows the steady-state flow strength $q=T-S$ as a function of the freshwater flux $\sigma$. We remark that for the parameters described above the Stommel box model (\ref{e.2Bnd1})--(\ref{e.2Bnd2}) is very close to the saddle-node. In Section~\ref{sec:numerics} we shall consider freshwater fluxes which allow for bistability with $\sigma=0.8$ and which support only a single stable solution with $\sigma=1.3$.



\subsection{Atmosphere model}
\label{sec:42}
We consider the Lorenz-84 model for the general circulation of the atmosphere \citep{Lorenz84,Lorenz90}
\begin{align}
\dot x &= -(y^2+z^2) - a(x-F) \nonumber\\
\dot y &= xy-bxz - (y-G)\nonumber \\
\dot z &=  bxy+xz- z  ,
\label{e.L84}
\end{align}
which evolves on the fastest time scale with typical times of the order of days. These equations describe the westerly zonal mean flow current with strength $x$ and the amplitudes $y,z$ of the cosine and sine waves of the mean circulation. The superimposed sine and cosine waves are advected by the mean flow, modelled here by the quadratic terms involving the factor $b$.  The model describes how energy vacillates between a zonal jet stream and a meandering jet stream. $F$ denotes the meridional temperature gradient and the model assumes that the zonal mean flow is in thermal balance, neglecting the effect of the eddies $(y,z)$. Similarly, $G$ denotes the longitudinal temperature gradient, i.e. the heating gradient between land and sea, which is driving $y$. The model exhibits chaos depending on the parameters $a$, $b$, $F$ and $G$. Reasonable time units in this model are $5$ days and $a<1$ and $b>1$ \citep{Lorenz90}. In Figure~\ref{fig:DOModel_sketch} the chaotic attractor is depicted for $F=8$, $G=1$, $a=0.25$ and $b=4$. For each hemisphere we assume that the dynamics is given by a Lorenz-84 system (\ref{e.L84}). The difference between the two hemispheres is in how far the ocean and the sea-ice couple into the atmospheric model via the meridional and zonal temperature gradients.  In the Southern hemisphere the effect of the Northern ocean and sea-ice is neglected and we assume constant temperature gradients with $F^{(S)}=F_0^{(S)}$ and $G^{(S)}=G_0^{(S)}$. In the Northern hemisphere, the ocean and the atmosphere are coupled and we follow \cite{Roebber95} to couple the Stommel box model (\ref{e.2Bnd1})--(\ref{e.2Bnd2}) with the Lorenz-84 model (\ref{e.L84}). The coupling of the fast atmosphere to the slow ocean occurs via the ambient atmospheric temperature gradient $\Theta$ and the freshwater influx $\sigma$. The backcoupling of the slow ocean to the fast atmosphere occurs via the meridional and zonal temperature gradients $F$ and $G$, respectively. We make the following assumptions (suppressing the superscript $N$ denoting the Northern hemisphere):
\begin{enumerate}[(i)]
\item The meridional temperature gradient $F$ in the Lorenz-84 model (\ref{e.L84}) is (in the absence of sea-ice) approximated by the meridional temperature gradient of the ocean $T=T_e-T_p$ with $F = F_0 + F_1 T$ with $F_1\ge 0$. 
\item The longitudinal gradient $G$ in the Lorenz-84 model (\ref{e.L84}) is dominated by the temperature difference of land and sea. Ignoring the diurnal cycle, we argue that near the equator the land heats up more than the ocean whereas in the polar region the ocean is warmer than the land (especially during winter). Hence, an increased oceanic meridional temperature gradient $T=T_e-T_p$ with warmer equatorial waters and colder polar waters, implies a decreased longitudinal temperature gradient decreases. Hence the longitudinal temperature gradient $G$ in the Lorenz-84 model (\ref{e.L84}) is (in the absence of sea-ice) modelled as  $G=G_0 - G_1  T$ with $G_1\ge 0$. 
\item The ambient temperature gradient $\Theta(t)=\Theta_e(t)-\Theta_p(t)$ in the Stommel box model (\ref{e.2Bnd1})--(\ref{e.2Bnd2}) is given by thermal wind balance as $\Theta = \theta x$ (in the absence of sea-ice) . Without sea-ice we would have $\Theta=(x-F_0)/F_1$.
\item The freshwater transport associated with evaporation and precipitation depends on the strength of the atmospheric eddies and we set $\sigma(t)=\sigma_0 + \sigma_1(y^2+z^2)$. Here $\sigma_0$ may be a function of time if freshwater fluxes stemming from melting glaciers is included. In this work, however, we do not consider any external freshwater flushes.
\end{enumerate}
Introducing the eddy strength $\Delta=y^2+z^2$, we summarise the ocean-atmosphere coupling as
\begin{align}
F  &= F_0 + F_1T \nonumber\\
G  &= G_0 - G_1 T \nonumber\\\
\Theta(t) & = \theta_0 + \theta_1\frac{x-\bar x}{\sqrt{\epsilon_f}}  \nonumber\\\
\sigma(t) &= \sigma_0 + \sigma_1\frac{\Delta-\bar\Delta}{\sqrt{\epsilon_f}} .
\label{eq:FGTS1}
\end{align}
Here and in the following a bar denotes the average. The atmospheric driving terms $(x-\bar x)/\sqrt{\epsilon_f}$ and $(\Delta-\bar\Delta)/\sqrt{\epsilon_f}$ converge to Brownian motion for $\epsilon_f\to 0$ as outlined in Section~\ref{sec:homo}. They represent the stochastic forcing of the atmosphere on the slow ocean dynamics.


\subsection{Sea-ice model}
\label{sec:43}
The presence of sea-ice significantly changes the dynamics of the slower ocean and the faster atmosphere. Sea-ice interacts with both the atmosphere and the ocean in several ways. Sea-ice responds rapidly to changes in temperature and grows on a typical time scale of a few months, placing its dynamics on an intermediate time scale between the fast atmospheric dynamics and the slow ocean dynamics. Sea-ice is created by colder polar ocean box temperatures $T_p$. Conversely, it is melted by warmer polar ocean temperatures $T_p$. Furthermore, the meridional atmospheric heat flux plays a major role in the melting and preservation of sea-ice \citep{MonahanEtAl08,DrijfhoutEtAl13,KleppinEtAl15}. In particular, meandering of the westerly Northern hemisphere jet stream enhances the meridional atmospheric heat flux by warm eddies drawing warm tropical air into polar regions. The degree of meandering of the jet stream is captured in our model by $\Delta_N=y_N^2+z_N^2$. \cite{BanderasEtAl12,BanderasEtAl14} showed that additionally Southern Ocean winds, measured in our model by the strength of the zonal mean flow $x_S$, couple the southern and northern oceans via Ekman pumping thereby influencing the sea-ice extent.  

We parametrise the sea-ice cover by a variable $\xi(t)$. We consider here intermittent sea-ice dynamics where the sea-ice cover exhibits sporadic brief periods of extreme extent. To model such dynamics we employ a CAM process (\ref{eq:CAM}). Acknowledging the atmospheric dynamics as a driver for the variations of sea-ice cover, we propose the following deterministic approximation of a CAM process,
\begin{align}
\epsilon_i \dot \xi &= (\lambda + \frac{\kappa^2}{2})\xi + \sqrt{\frac{\epsilon_i}{\epsilon_f}}\delta (\kappa\xi+g)(x_S-\bar x_S)+ \sqrt{\frac{\epsilon_i}{\epsilon_f}}c\, (\Delta_N-\bar\Delta_N),
\label{eq:CAM_MS}
\end{align}
where the noise is deterministically generated by the chaotic atmospheric northern eddies $\Delta_N(t)$ and the effect of the southern zonal jet stream $x_S(t)$. We assume for simplicity that this effect scales linearly with $\Delta_N(t)$ and $x_S(t)$, respectively. According to the theory of deterministic homogenisation presented in Section~\ref{sec:homo}, this ordinary differential equation converges for $\epsilon_f\to 0$, i.e. when the atmosphere is infinitely faster than the sea-ice dynamics, to the CAM stochastic differential equation
\begin{align}
\epsilon_i d\xi &= (\lambda + \kappa^2)\xi \, dt +(\kappa \xi + g) \circ dW_1 + \tilde c \, dW_2.
\label{eq:CAMstoch}
\end{align}
The limiting stochastic differential equation (\ref{eq:CAMstoch}) corresponds to the CAM process (\ref{eq:CAM}) with $L=\lambda+\kappa^2/2$, $E=\delta \eta_x\kappa$, $B=\tilde c =\eta_\Delta c$ and $G=\delta \eta_x g/(1+E^2/(2L))$ and with $y_\eps=\xi-A$ where $A=EG/(2L)$. Here $\eta_{x,\Delta}$ are the standard deviations of the noises $W_x(t)=\lim_{{\epsilon_f}\to 0}\int^{t/\epsilon_f}(x_S(s)-\bar x_S) ds/\sqrt{\epsilon_f}$ and $W_\Delta(t)=\lim_{{\epsilon_f}\to 0}\int^{t/\epsilon_f}(\Delta_N(s)-\bar \Delta_N) ds/\sqrt{\epsilon_f}$. Note that whereas actual sea-ice cover is a bounded variable, the variable $\xi(t)$ is unbounded. In this sense the CAM process (\ref{eq:CAM_MS}) (and its limiting dynamics (\ref{eq:CAMstoch})) does not model the actual extent of the sea-ice but rather constitutes a conceptual model to account for the assumed intermittent nature of the sea-ice cover.\\

The influence of sea-ice on the ocean and atmosphere is manifold. Sea-ice acts as a thermal insulator, preventing the exchange of heat from the ocean to the atmosphere, thereby decreasing the meridional ocean temperature gradient $T=T_e-T_p$. This effect plays a major role in our model and will be shown to be responsible for the abrupt temperature changes. Once sea-ice has formed it prohibits precipitation of evaporated water from the polar ocean on polar land mass, suppressing freshwater fluxes. Furthermore, during the formation of sea-ice salt is extruded into the ocean during build up and freshwater is added into the ocean during melting. 
Sea-ice affects both meridional and longitudinal temperature gradients of the atmosphere (i.e. $F^{(N)}$ and $G^{(N)}$ in our model). Increased sea-ice extent strengthens the meridional thermal gradient experienced by the atmosphere, thereby increasing the zonal mean-flow component $x_N$. Similarly, an increased sea-ice extent leads to a decreased longitudinal thermal gradient experienced by the atmosphere, thereby decreasing $G$ (again favouring zonal flow $x_N$). This motivates to augment the expressions for the meridional and longitudinal temperature gradients of the atmosphere $F$ and $G$ in the Lorenz-84 model (\ref{e.L84}) (for the Northern hemisphere) and the ambient oceanic temperature gradient $\Theta$ and the freshwater flux $\sigma$ in the Stommel box model (\ref{e.2Bnd1})--(\ref{e.2Bnd2}). In particular we note (suppressing the superscript $N$):
\begin{enumerate}[(i)]
\item The meridional thermal gradient in the Northern hemisphere is given by the ocean temperature gradient $T$ if there is no sea-ice ($\xi=0$) and is increased by sea-ice $\xi>0$ independent of the ocean temperature gradient: 
\begin{align}
F &= F_0 + F_1T + F_2\xi ,
\label{eq:SI_1}
\end{align}
with $F_{1,2}\ge 0$. Note that in the case of sea-ice $\xi>0$, the equatorial sea temperature $T_e$ continues to contribute to the thermal gradient, so the oceanic temperature gradient $T$ is still affecting $F$ with $F_1\neq 0$ even in the presence of sea-ice.
\item The longitudinal thermal gradient  in the Northern hemisphere is dominated by the ocean temperature gradient $T$ if there is no sea-ice ($\xi=0$) and is decreased by sea-ice $\xi>0$ independent of the ocean temperature gradient
\begin{align}
G = G_0 - G_1 T - G_2\xi , 
\label{eq:SI_2}
\end{align}
with $G_{1,2}\ge 0$. As for the meridional thermal gradient discussed above in (i), the land-sea temperature gradient at the equator is still determined by the equatorial ocean temperature $T_e$, so the oceanic temperature gradient $T$ is still affecting $G$ with $G_1\neq 0$ even in the presence of sea-ice.
\item The atmospheric temperature gradient  $\Theta(t)=\theta x$ is maintained by thermal balance, so only indirectly affected by sea-ice. To account for the insulating effect of sea-ice a damping term proportional to $(\xi-\bar \xi)T$, where $\bar\xi$ denotes the mean of the sea-ice cover variable $\xi$,  is added to the temperature gradient equation (\ref{e.2Bnd1}). This term  (cf. (\ref{eq:ocean})) is the key dynamical ingredient for the generation of abrupt sharp temperature changes in our model, resembling DO events. To highlight the role of the intermittent sea-ice events we introduce a thresholded driver $\Xi(\xi)={\rm{max}}(\xi,\xi^\star)$ which filters out small fluctuations with $\xi<\xi^\star$. We shall use this thresholded driver, upon subtracting its mean $\bar\Xi$, to enter the ocean dynamics and consider a damping term of the form $(\Xi(t)-\bar\Xi)T$ in the temperature gradient equation (\ref{e.2Bnd1}). 
\item The source term of salinity decreases during growth of sea-ice and increases during melting of sea-ice. We set 
\begin{align}
\sigma(t) = \sigma_0 + \sigma_1(y^2+z^2) - \sigma_2 \dot \xi .
\label{eq:SI_3}
\end{align}
\end{enumerate}
Summarising we motivated the proposed coupled ocean-atmosphere and sea-ice model (\ref{eq:ocean})--(\ref{eq:atmosphere}) with the interactions captured in (\ref{eq:FG})--(\ref{eq:sigma}), which are expressed by (\ref{eq:SI_1})--(\ref{eq:SI_3}). In the next section we will illustrate how this model is able to reproduce abrupt temperature changes as in DO events.


\section{Illustration of the model}
\label{sec:numerics}

We now show numerical simulations of the conceptual coupled ocean-atmosphere and sea-ice model (\ref{eq:ocean})--(\ref{eq:atmosphere}). We focus here on the effect of intermittent sea-ice on the oceanic temperature gradient $T$ through insulation, as expressed by the linear damping term in (\ref{eq:ocean}). 

In the Stommel box model we set $\mu=7.5$ and set $\epsilon_a=0.34$, corresponding to the relaxation time in the deep ocean (we have checked that our results do not depend qualitatively when varying $\epsilon_a$). We choose as base ambient temperature gradient $\theta_0=1$ and as base freshwater flux we consider here $\sigma_0=0.8$ for which the uncoupled Stommel box model exhibits bistability and $\sigma_0=1.3$ for which only a single stable solution exists (cf. Figure~\ref{fig:Stommel_q}). The perturbations to these base states induced by atmospheric noise are set to $\theta_1=0.01/\eta_x$ and $\sigma_1=0.01/\eta_\Delta$ and neglect the effect of sea-ice on the freshwater flux setting $\sigma_2=0$. We further suppress the backcoupling of the slow ocean dynamics onto the fast atmospheric dynamics by setting $F_1=G_1=0$. The standard deviations of the atmospheric noise associated with zonal mean flow $x$ and the large-scale eddies $\Delta$, respectively, $\eta_x=0.513$ and $\eta_\Delta=0.071$, were estimated from a long time-integration of the Lorenz-84 model. The atmosphere is kept in perpetual winter conditions with $F_0=8$ and $G_0=1$ and with $a=0.25$ and $b=4$ \citep{Lorenz84}. We choose for simplicity the same values of the parameters $a,b,F_0,G_0$ for the Northern and the Southern hemisphere. This is not necessary; the only requirement in the derivation of the deterministic approximation of the CAM noise model for sea-ice is that the northern and southern atmospheric dynamics are sufficiently decorrelated which can be achieved using the same equation parameters but different initial conditions. The sea-ice is coupled to the Stommel two-box model with $d=50$, and its parameters are set to $\kappa=1.118$, $\lambda=-1.565$, $g=0.3351$, $\delta=1/\sigma_1$ and $c=0.3/\eta_2$. Similarly the mean values $\bar x = 1.0147$, $\bar\Delta = 1.7463$ and $\bar\xi=0.12$ were estimated from long time simulations of the Lorenz-84 model and the sea-ice model. Note that in the limit $\epsilon_f\to 0$ we expect $\bar \xi=0$. The physical set-up suggests that in the Stommel box model a unit of time corresponds to $219$ years, and that the time-scale parameters controlling the time-scales of the fastest atmospheric processes and the intermediate time scale of the sea-ice are $\epsilon_f=0.0083$ and $\epsilon_i=0.05$ (cf. (\ref{eq:eps})).\\

\begin{figure}
	\centering
	\includegraphics[width=0.47\linewidth]{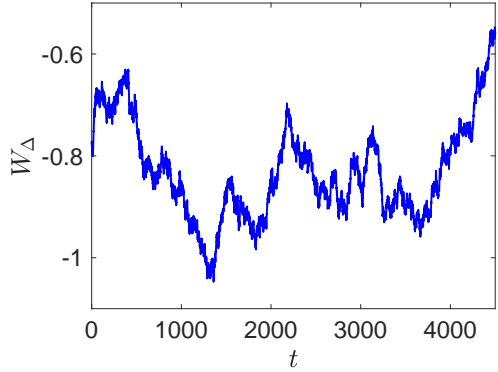}		
	\caption{Time series of $W_\Delta$ (\ref{eq:DeltaNoise}) approximating Gaussian noise.}
	\label{fig:HDelta}
\end{figure}

\begin{figure}
	\centering
	\includegraphics[width=0.47\linewidth]{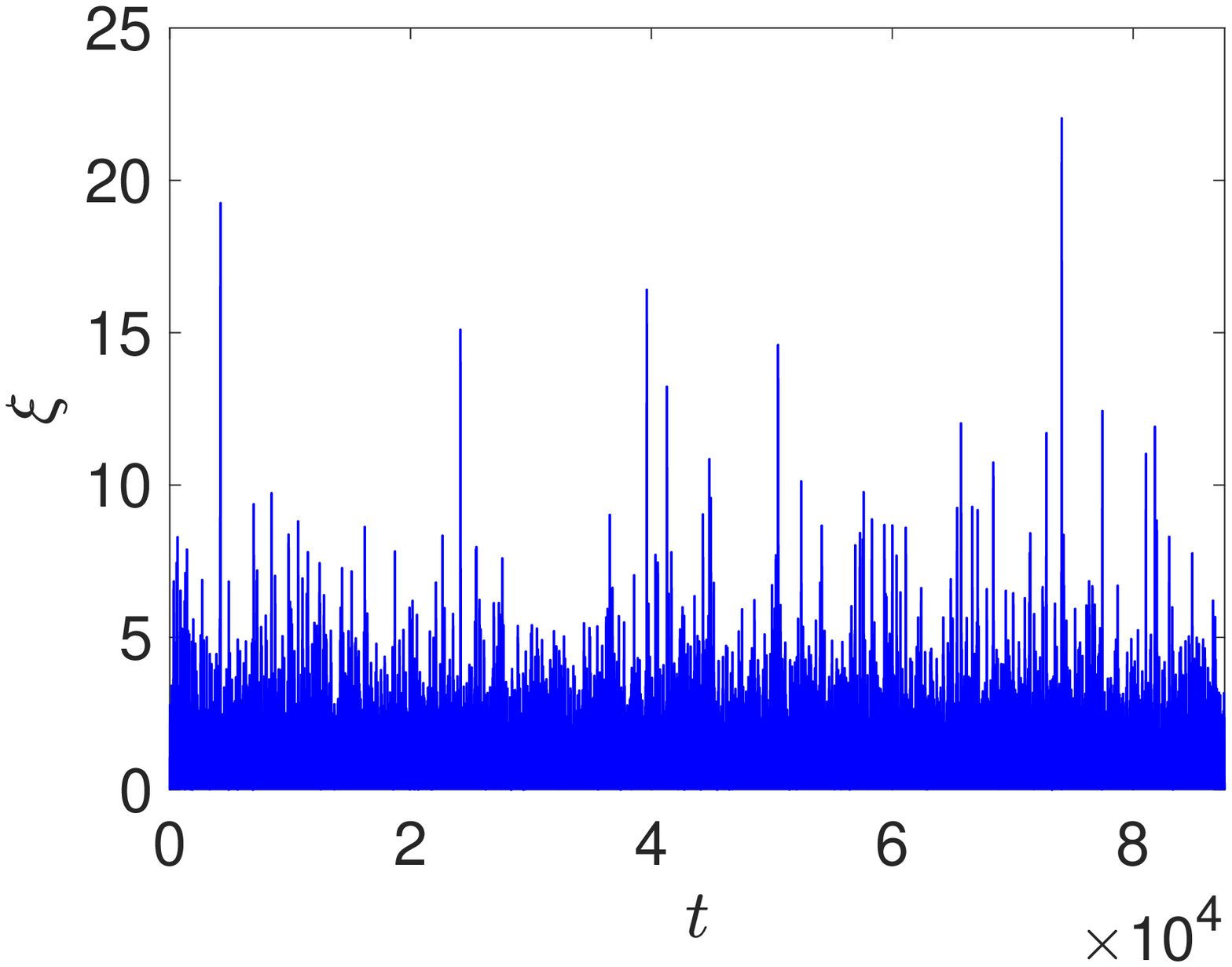}		
	\hspace{1mm}
	\includegraphics[width=0.47\linewidth]{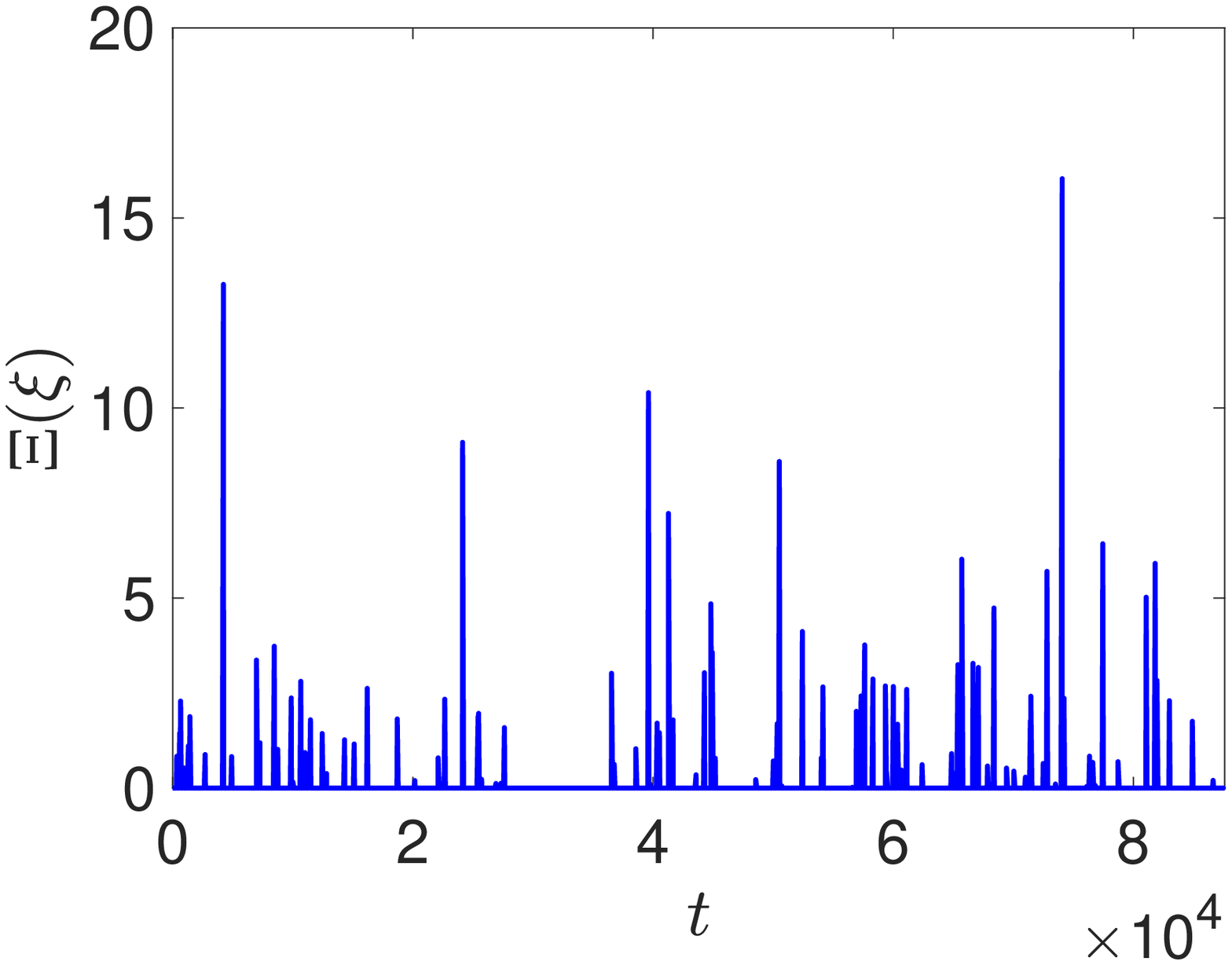}	
	\caption{Left: Time series of the sea-ice variable $\xi(t)$ approximating CAM noise. Right: Time series of the associated threshold time series $\Xi(\xi)={\rm{max}}(\xi,6)-6$.}
	\label{fig:xi}
\end{figure}

\begin{figure}
	\centering
	\includegraphics[width=0.47\linewidth]{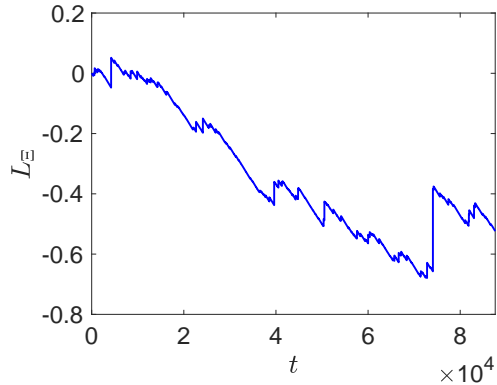}	
	\caption{Integrated noise $L_\Xi$ (\ref{eq:xiNoise}) approximating and $\alpha$-stable process with $\alpha=1.5$ and $\beta=0.99$ for the time series $\Xi(\xi)$ depicted in Figure~\ref{fig:xi}.}
	\label{fig:Hxi}
\end{figure}

We first illustrate the various statistical limit laws which give rise to the effective stochastic behaviour of the deterministic coupled ocean-atmosphere and sea-ice model (\ref{eq:ocean})--(\ref{eq:atmosphere}). We confirm the deterministic approximation of stochastic Gaussian processes $W_t$ by
\begin{align}
W_x(t) &= \frac{1}{\sqrt{\epsilon_f}}\int^{\frac{t}{\epsilon_f}}(x(s)-\bar x)ds
\label{eq:xNoise}
\\
W_\Delta(t)& = \frac{1}{\sqrt{\epsilon_f}}\int^{\frac{t}{\epsilon_f}}(\Delta(s)-\bar\Delta)ds,
\label{eq:DeltaNoise}
\end{align}
and of the L\'evy processes $L_{\alpha,\eta,\beta}$ by
\begin{align}
L_\xi(t) = \frac{1}{\epsilon_i^{1-\gamma}}\int^{\frac{t}{\epsilon_i}}(\xi(s)-\bar \xi)ds,
\label{eq:xiNoise}
\end{align}
with $\gamma=1/\alpha$. These constitute the noise processes driving the coupled model (\ref{eq:ocean})--(\ref{eq:atmosphere}). We show results in Figure~\ref{fig:HDelta} for the approximation of Gaussian noise $W_\Delta$ (plots for $W_x$ look similar). Figure~\ref{fig:xi} shows a realisation of the time series of the sea-ice variable $\xi(t)$ obtained from (\ref{eq:seaice}), as well as the thresholded driver $\Xi(\xi)$ which captures the intermittent large sea-ice cover events above the threshold $\xi^\star=6$. The corresponding integrated noise approximation $L_\Xi$ is shown in Figure~\ref{fig:Hxi}. The parameters chosen for the sea-ice model (\ref{eq:seaice}) imply $\alpha=1.5$ and $\beta=0.99$ (cf. (\ref{eq:alphaCAM}) and (\ref{eq:betaCAM})). The integrated CAM-process $L_\xi$ and the thresholded version $L_\Xi$ exhibit almost exclusively positive jumps as predicted by the homogenisation theory results which yields $\beta=0.99$.\\

\begin{figure}
	\centering
          \includegraphics[width=0.47\linewidth]{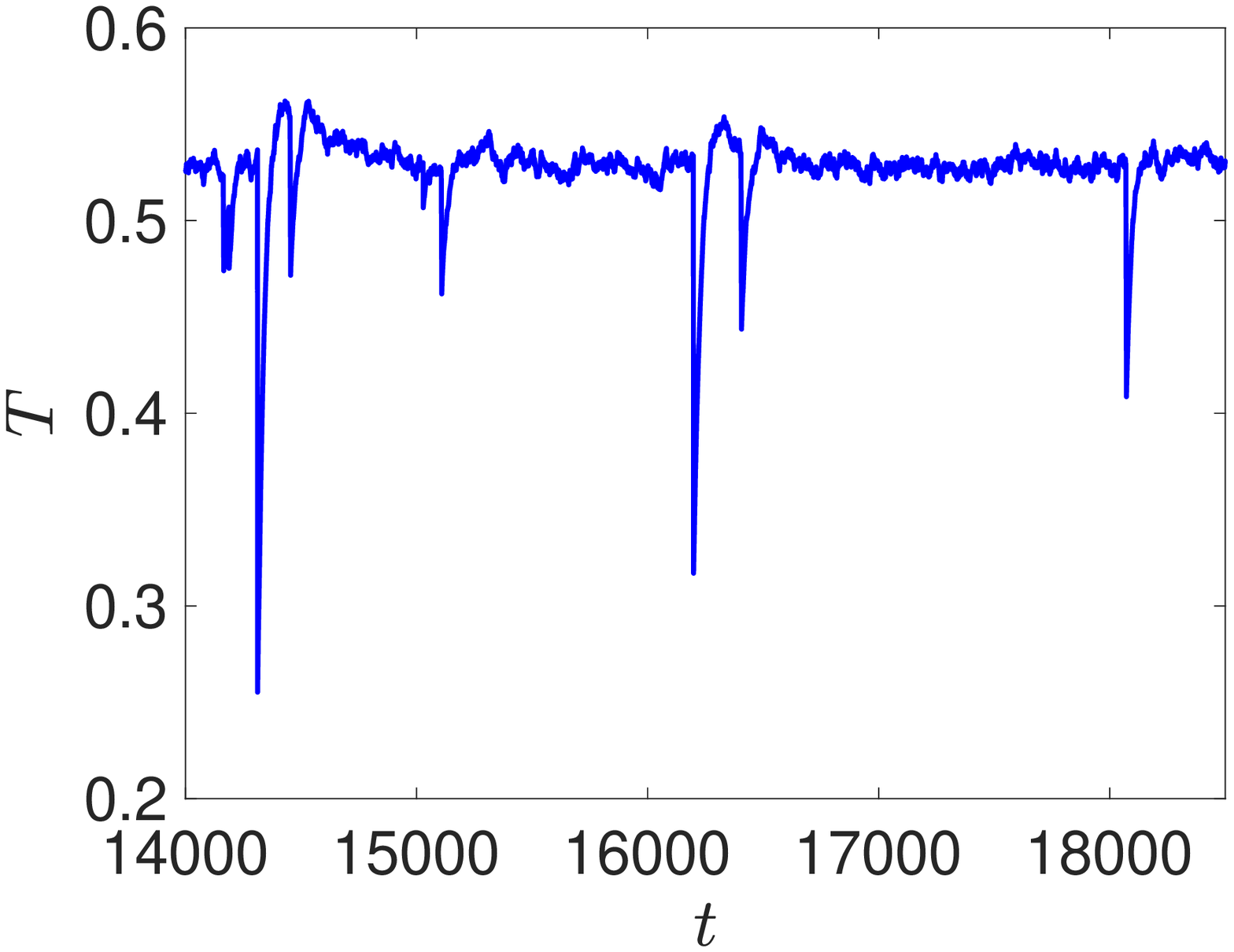}\;		
	\includegraphics[width=0.47\linewidth]{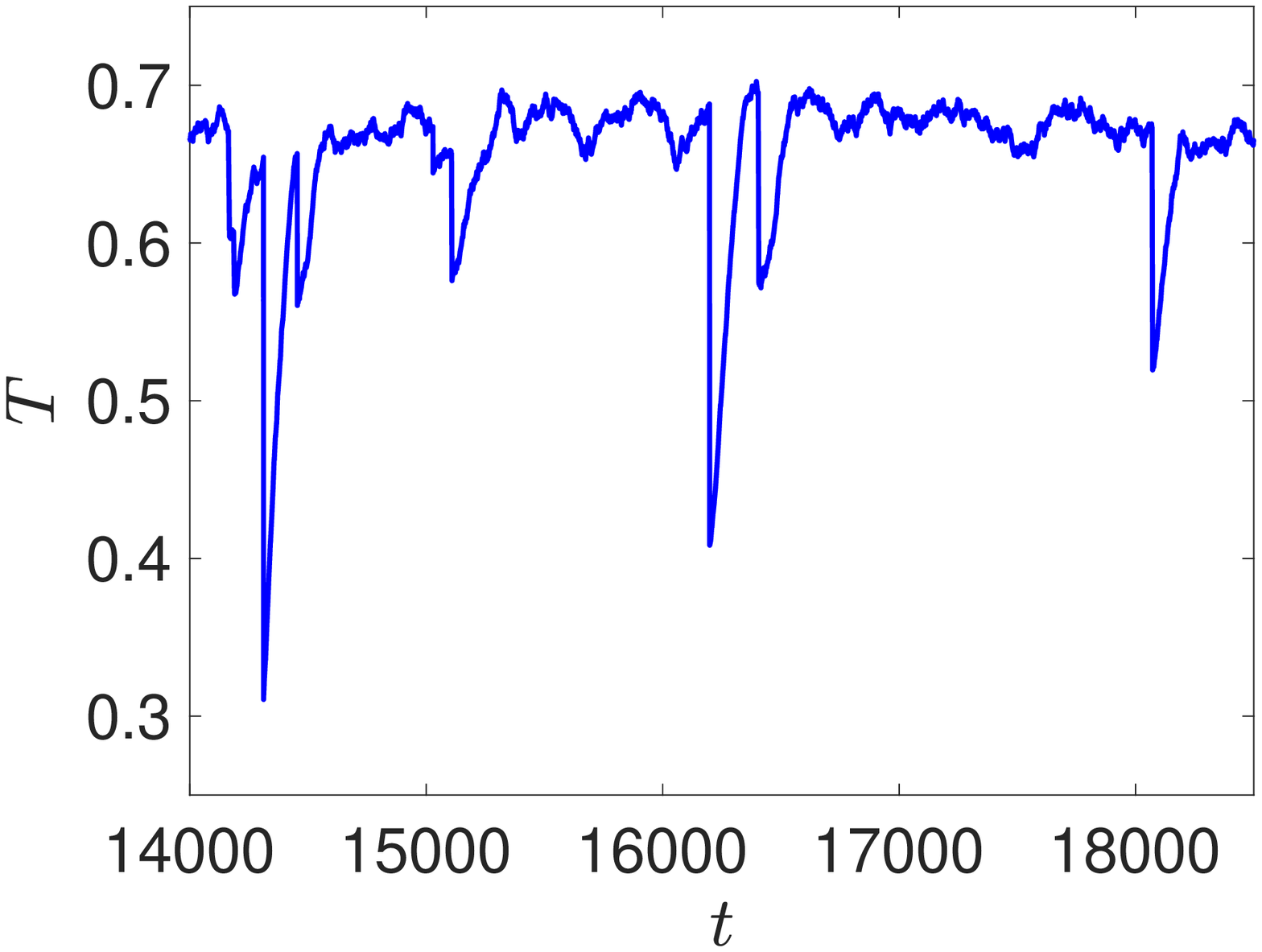}
	\caption{Time-series of the oceanic meridional temperature gradient $T$ obtained by integration of the model (\ref{eq:ocean}) driven by the sea-ice time-series depicted in Figure~\ref{fig:xi}. Left: $\sigma_0=0.8$. Right: $\sigma_0=1.3$.}
	\label{fig:T}
\end{figure}

\begin{figure}
	\centering
	\includegraphics[width=0.47\linewidth]{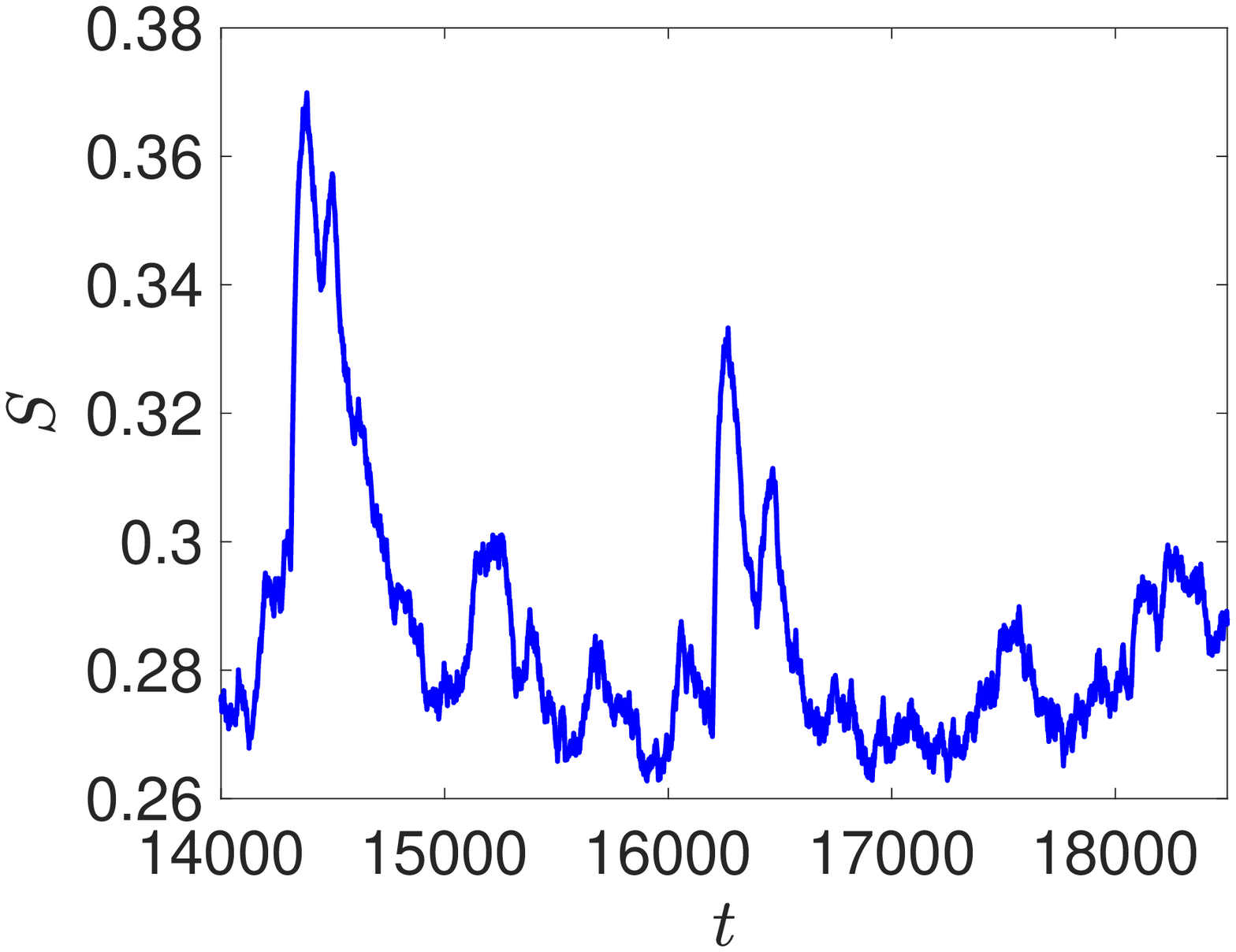}\;	
	\includegraphics[width=0.47\linewidth]{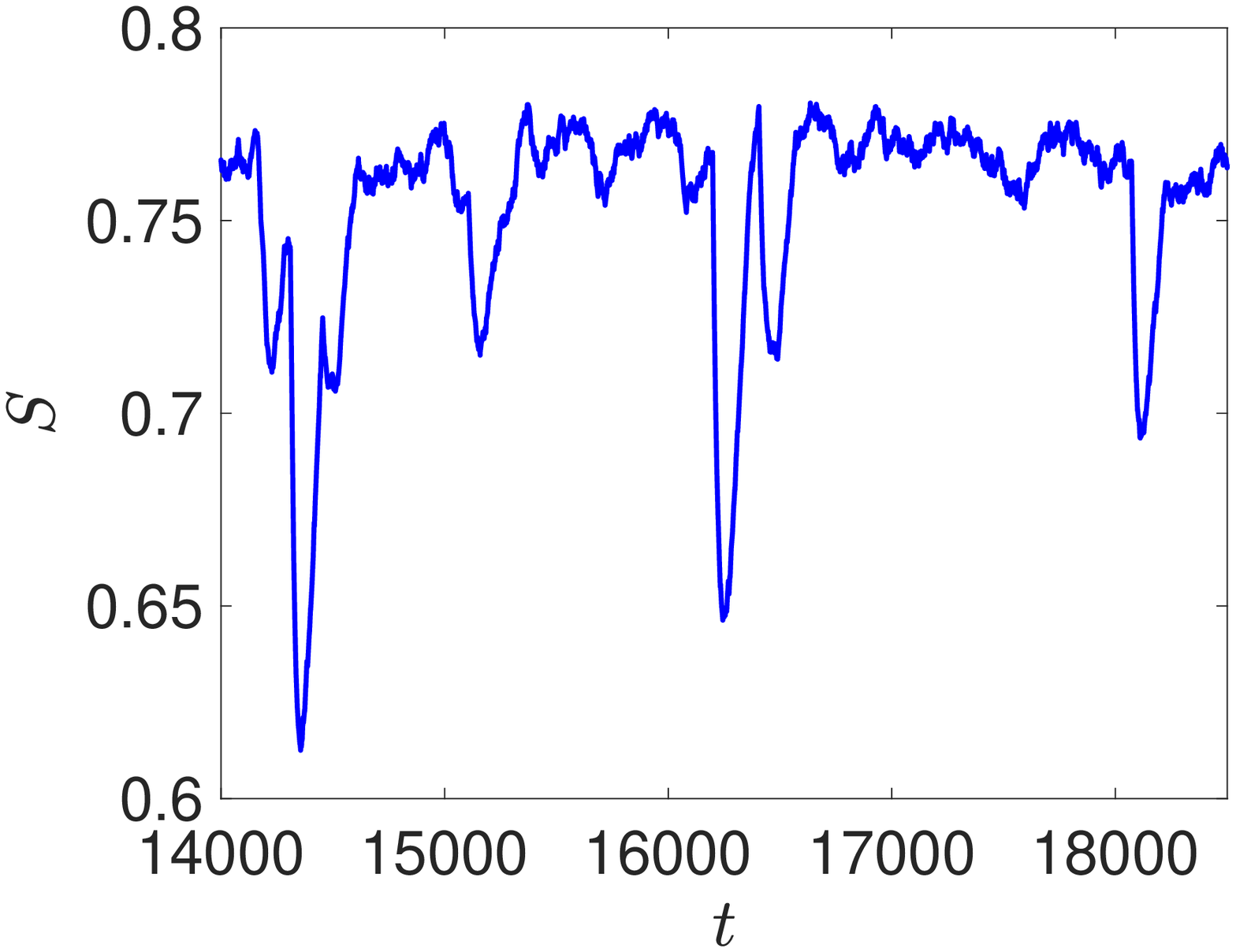}	
	\caption{Time-series of the salinity $S$ obtained by integration of the model (\ref{eq:ocean}) driven by the sea-ice time-series depicted in Figure~\ref{fig:xi}. Left: $\sigma_0=0.8$. Right: $\sigma_0=1.3$.}
	\label{fig:S}
\end{figure}

\begin{figure}
	\centering
	\includegraphics[width=0.47\linewidth]{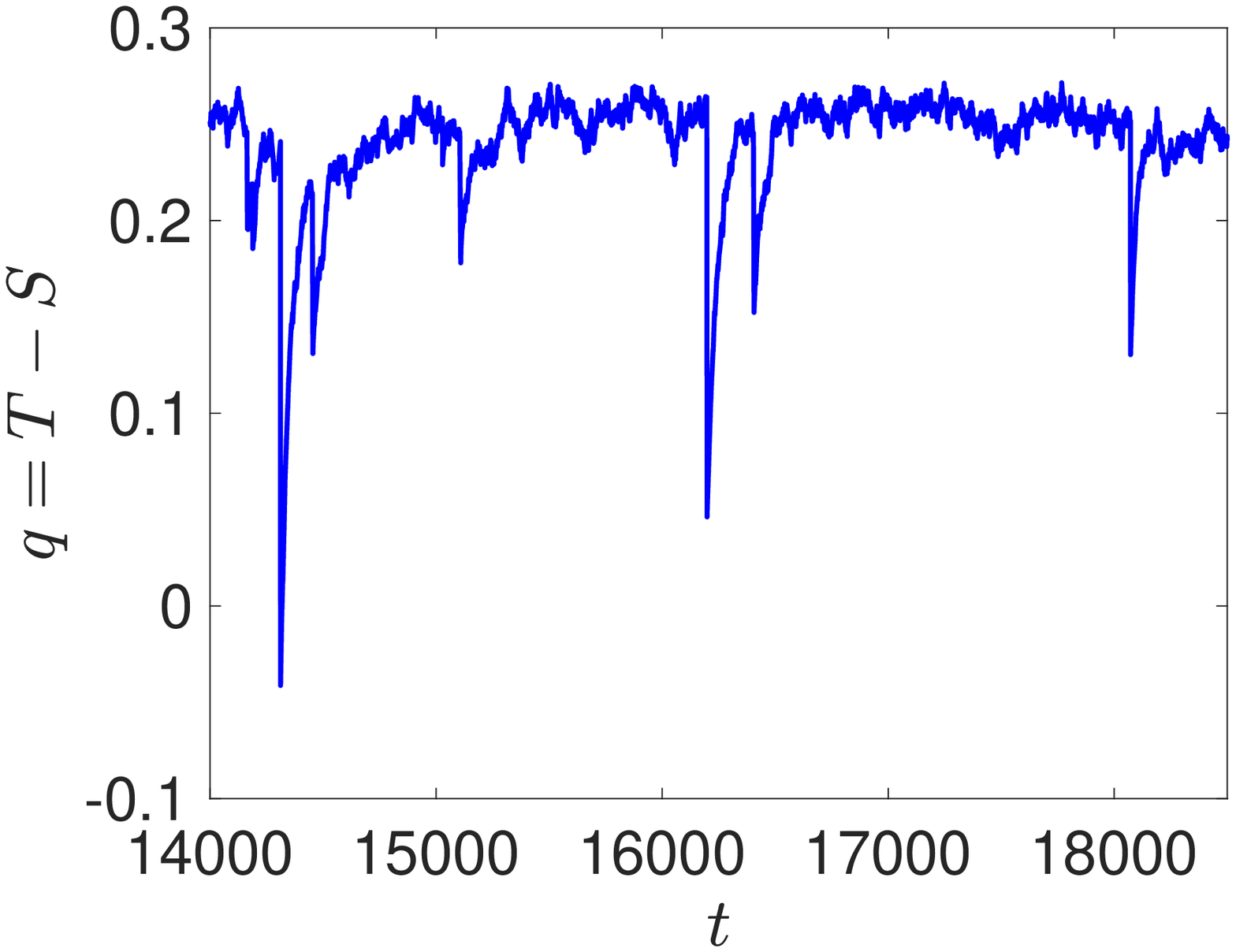}\;	
	\includegraphics[width=0.47\linewidth]{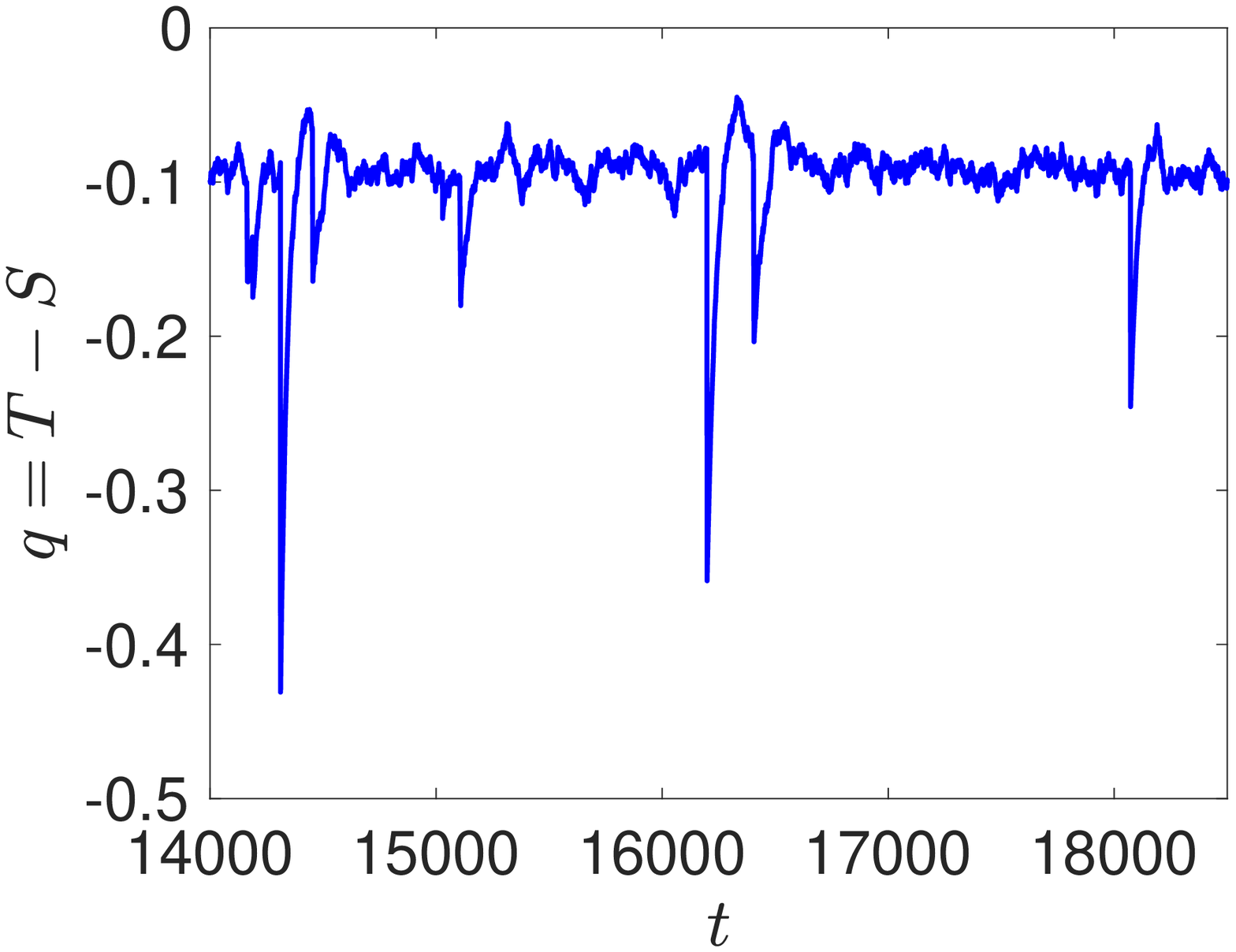}	
	\caption{Time-series of the flux $q=T-S$ obtained by integration of the model (\ref{eq:ocean}) driven by the sea-ice time-series depicted in Figure~\ref{fig:xi}. Left: $\sigma_0=0.8$. Right: $\sigma_0=1.3$.}
	\label{fig:q}
\end{figure}

The effect of these jumps on the ocean's temperature gradient $T$ is illustrated in Figure~\ref{fig:T} where we show results for $\sigma_0=0.8$ and for $\sigma_0=1.3$. For $\sigma_0=0.8$ the uncoupled Stommel box model supports two stable solutions, and the abrupt changes are shown as deviations of the interstadial solution which is characterised by a positive thermally-driven flux $q=T-S>0$. For $\sigma_0=1.3$ the Stommel box model only supports a single solution which is characterised by negative salinity-driven flux $q<0$. In both cases, the $\alpha$-stable driver $L_\Xi$ leads to significant sharp drops on the meridional temperature gradient $T=T_e-T_p$, implying  sharp increases of the oceanic polar temperature $T_p$. At $t\approx 14,300$ this is particularly strong with a change in temperature of more than $7^\degree$C (the Stommel model is normalised such that $T=1$ corresponds to $20^\degree$C). This large and abrupt change is caused by the large jump of $L_\xi$ which itself is caused by a prolonged period of large sea-ice cover events $\xi$ (cf. Figure~\ref{fig:Hxi}). These temperature increases gradually decay to the (noisy) steady interstadial state. The time between events is here roughly $1,800$ years, which is the same order of magnitude as observed in ice-core records. The corresponding time-series for the salinity $S(t)$ and the flux $q(t)=T-S$ are shown in Figure~\ref{fig:S} and Figure~\ref{fig:q}. Whereas the salinity gradient increases for $\sigma_0=0.8$ it decreases for $\sigma_0=1.3$ during the abrupt changes. In both cases, the resulting flux $q$ decreases, implying a more salinity-driven transport during the abrupt changes.\\ 

An application of the $p$-variation test, described in Section~\ref{sec:Data}, determines the stability parameter of the time-series for the meridional temperature gradient $T$ as $\alpha=1.8$ for $\sigma_0=0.8$ and $\alpha=1.75$ for $\sigma_0=1.3$, consistent with the value of $\alpha=1.78$ obtained in Section~\ref{sec:Data} from $\ch{Ca}^{2+}$  ice-core data and the results by \cite{Ditlevsen99}. The small fluctuations of $T$ and $S$ are induced by fast atmospheric (Brownian) noise with $\theta_1\neq 0$ and $\sigma-1\neq 0$, respectively.\\






\section{Discussion}
\label{sec:discussion}

We developed a self-consistent conceptual model of a slow ocean coupled to a fast atmosphere and to sea-ice, which evolves on an intermediate time scale and is driven by the atmosphere. The model relates the abrupt climate changes of DO events to intermittent sea-ice dynamics and the sporadic occurrence of large sea-ice extent. The intermittency in the sea-ice model is induced by synergetic forcing by fast atmospheric Northern hemisphere eddy activity and by fast atmospheric Southern hemisphere zonal mean flow. The sea-ice then acts on the slow ocean by insulating it, preventing the heat exchange of the polar ocean with the atmosphere. Using statistical limit laws for deterministic chaotic dynamical systems the sea-ice model was shown to generate non-Gaussian $\alpha$-stable noise, consistent with the time series analysis of ice core data \citep{Ditlevsen99}. The apparent regularity of the temporal spacing between successive Dansgaard-Oeschger events deduced from the ice-core data \citep{GrootesStuiver97,YiouEtAl97,DitlevsenEtAl05}, is here not caused by any inherent periodicity in the system but rather by the random occurrence of extreme sea-ice extents above a certain threshold below which the response of the ocean is not significant. This is in accordance with \cite{DitlevsenEtAl07} who showed that there is no statistically significant evidence for strict periodicity.\\

The particular signature of the temperature with its abrupt warming events is caused by an intermittent process evolving on a faster time scale than the oceanic time scale. In our model here this process is provided by (the approximation of) a CAM process $\xi$ (cf (\ref{eq:CAM_MS})) which quantifies the variability in the sea-ice cover. The integrated CAM noise in the variable the gives rise to non-Gaussian $\alpha$-stable statistics with the jumps corresponding to the abrupt warming events. The CAM noise itself was dynamically induced by fast atmospheric noise. It is pertinent to mention that one could equally consider other intermittent mechanisms than sea-ice cover variability such as intermittent freshwater influxes. In this case the CAM noise would enter the salinity equation (\ref{eq:ocean}) via the freshwater source terms in $\sigma(t)$ (\ref{eq:sigma}), and the CAM noise would be a conceptual model for intermittent freshwater changes, captured by $\dot \xi$.\\

The model hinges on statistical limit laws. These laws were invoked to generate both the Brownian noise as well as the non-Gaussian $\alpha$-stable noise. Statistical limit laws describe the statistical properties of integrals (or sums) of observables. The observables here are observables of (relatively) fast variables. The integrals over the observables naturally arise in the multi-scale context when the faster variables are integrated in the slower dynamics. The simplest statistical limit law is the law of large numbers, which ensures that appropriately scaled variables (here our observables) converge to a deterministic limit, their average. The central limit theorem and its generalisations allows precise statements on fluctuations around the mean behaviour. Whereas statistical limit laws are part of the standard tool box when the observations are of a stochastic nature, and in particular when the observations are independent identically distributed random variables. The case of integrals (or sums ) of deterministic chaotic observables has only been recently explored. These studies provide a rigorous justification why scientists can parametrise the effect of unresolved scales, such as the effect of fast weather on the slow ocean, by noise as proclaimed by \cite{Hasselmann76} and \cite{Leith75} in the context of climate dynamics. Rather than just providing a general qualitative framework, statistical limit theorems and homogenisation theory provide precise statements on the nature of the noise -- i.e. is the noise Brownian or $\alpha$-stable, is it additive or multiplicative, and is the noise to be interpreted in the sense of It\^o or of Stratonovich/Marcus? Furthermore, homogenisation theory provides explicit expressions for the drift and diffusion coefficients of the limiting stochastic differential equation. Recently, at least formally, statistical limit laws were extended to the more realistic case of finite time-scale separation \citep{WoutersGottwald19a,WoutersGottwald19b}. The typical application of statistical limit laws in the geosciences is to provide closed equations for resolved variables of interest by parametrising unresolved fast and/or small-scale degrees of freedom by noise. The reward for such a parametrisation is of a computational nature as one now only needs to simulate an equation on the slow time scale, avoiding prohibitively small time steps needed to control numerical instabilities of the fast dynamics. 

Here we pursue a conceptionally different route. Rather than starting from a deterministic dynamical system to derive a limiting stochastic dynamical system, we reverse the order and use statistical limit laws to determine dynamical mechanisms which are consistent with the statistical properties of the observations. We use statistical limit laws in the sense of reverse engineering, thereby identifying key dynamical mechanisms for DO events such as intermittency,  provided by sea-ice variability on an intermediate time-scale. Statistical limit laws allowed us to both generate the intermittent process in the first place (here we used atmospheric noise to generate the intermittent CAM process for the sea-ice dynamics) as well as generating the $\alpha$-stable process driving the slow ocean dynamics with its abrupt climate changes. The former was achieved by central limit theorems generating Brownian motion, the latter by a generalised central limit theorem generating non-Gaussian L\'evy processes.

%
%



\section*{Acknowledgments}
The ice core data were generously provided by Peter Ditlevsen. I am grateful to  Armin K\"ohl, Johannes Lohmann, Marisa Montoya and Xu Zhang for many interesting and helpful discussions. I would like to thank Cameron Duncan, Nathan Duingan and Eric Huang who explored the $p$-variation test and suitable parameter ranges of the Lorenz-84 system in a summer project in 2014 at an early stage of this work. 


\bibliographystyle{natbib}




\end{document}